\documentclass[american,aps,pra,reprint,superscriptaddress, longbibliography]{revtex4-1}
\usepackage[T1]{fontenc}
\usepackage[latin9]{inputenc}
\usepackage{color}
\usepackage{babel}
\usepackage{amsthm}
\usepackage{amsmath}
\usepackage{amssymb}
\usepackage{graphicx}
\usepackage[unicode=true,pdfusetitle, bookmarks=true,bookmarksnumbered=false,bookmarksopen=false,  breaklinks=false,pdfborder={0 0 0},backref=false,colorlinks=false] {hyperref}
\hypersetup{ colorlinks,linkcolor=myurlcolor,citecolor=myurlcolor,urlcolor=myurlcolor}
\usepackage{tcolorbox}

\makeatletter
\@ifundefined{textcolor}{}
{%
 \definecolor{BLACK}{gray}{0}
 \definecolor{WHITE}{gray}{1}
 \definecolor{RED}{rgb}{1,0,0}
 \definecolor{GREEN}{rgb}{0,1,0}
 \definecolor{BLUE}{rgb}{0,0,1}
 \definecolor{CYAN}{cmyk}{1,0,0,0}
 \definecolor{MAGENTA}{cmyk}{0,1,0,0}
 \definecolor{YELLOW}{cmyk}{0,0,1,0}
 }
\theoremstyle{plain}

\theoremstyle{plain}

\ifx\proof\undefined
\newenvironment{proof}[1][\protect\proofname]{\par
\normalfont\topsep6\p@\@plus6\p@\relax
\trivlist
\itemindent\parindent
\item[\hskip\labelsep
\scshape
#1]\ignorespaces
}{%
\endtrivlist\@endpefalse
}
\providecommand{\proofname}{Proof}
\fi
\theoremstyle{plain}

\providecommand{\lemmaname}{Lemma}
\providecommand{\definitionname}{Definition}
\providecommand{\propositionname}{Proposition}

\usepackage{babel}
\usepackage{txfonts}
\usepackage{colortbl}\definecolor{myurlcolor}{rgb}{0,0,0.7}
\newcommand{\tr}{{\operatorname{Tr\,}}}

\def\ket#1{| #1 \rangle}
\def\bra#1{\langle  #1 |}
\def\braket#1{\langle  #1 \rangle}

\newcommand{\ketbra}[2]{|#1\rangle\!\langle#2|}

\newcommand{\haH}

\newcommand{\norm}[1]{\left\| #1 \right\|}

\definecolor{orange}{RGB}{255,127,0}

\newcommand{\aop}{\hat{a}_1}
\newcommand{\adop}{\hat{a}_1^\dag}
\newcommand{\aaop}{\hat{a}_2}
\newcommand{\aadop}{\hat{a}_2^\dag}
\newcommand{\mean}[1]{\left\langle{#1}\right\rangle}

\begin{document}
\title{Quantum Heat Engines with Carnot Efficiency at Maximum Power}

\author{Mohit Lal Bera}
\affiliation{ICFO -- Institut de Ci\`encies Fot\`oniques, The Barcelona Institute of Science and Technology, ES-08860 Castelldefels, Spain}

\author{Sergi Juli\`a-Farr\'e}
\affiliation{ICFO -- Institut de Ci\`encies Fot\`oniques, The Barcelona Institute of Science and Technology, ES-08860 Castelldefels, Spain}

\author{Maciej Lewenstein} 
\affiliation{ICFO -- Institut de Ci\`encies Fot\`oniques, The Barcelona Institute of Science and Technology, ES-08860 Castelldefels, Spain}
\affiliation{ICREA, Pg.~Lluis Companys 23, ES-08010 Barcelona, Spain} 

\author{Manabendra Nath Bera}
\email{mnbera@gmail.com}
\affiliation{Department of Physical Sciences, Indian Institute of Science Education and Research (IISER), Mohali, Punjab 140306, India}

\begin{abstract}
Heat engines constitute the major building blocks of modern technologies. However, conventional heat engines with higher power yield lesser efficiency and vice versa and respect various power-efficiency trade-off relations. This is also assumed to be true for the engines operating in the quantum regime. Here we show that these relations are not fundamental. We introduce quantum heat engines that deliver maximum power with Carnot efficiency in the \emph{one-shot finite-size} regime. These engines are composed of working systems with a finite number of quantum particles and are restricted to one-shot measurements. The engines operate in a one-step cycle by letting the working system simultaneously interact with hot and cold baths via semi-local thermal operations. By allowing quantum entanglement between its constituents and, thereby, a coherent transfer of heat from hot to cold baths, the engine implements the fastest possible reversible state transformation in each cycle, resulting in maximum power and Carnot efficiency. Finally, we propose a physically realizable engine using quantum optical systems. 
\end{abstract}

\maketitle

\section{Introduction}
Since the beginning of the industrial revolution, heat engines have been playing pivotal roles in shaping modern technologies. One of the central laws of thermodynamics for heat engines in the classical regime, that is, the second law, imposes a  fundamental limit on the maximum heat-to-work conversion efficiency in an engine, given by Carnot efficiency. This efficiency is only achieved when the engine operates in a cycle using reversible transformations, which requires it to run infinitely slowly. As a consequence, the engine's power - work extracted per unit time - becomes close to null. In general, the realistic engines operate in finite time to deliver a non-vanishing power, and then, the efficiency is compromised. The trade-off between efficiency and power is studied extensively in the past decades; see, for example, \cite{Curzon75, Berry00, Salamon01}, in the context of finite-time classical engines.   

In general, the laws of thermodynamics cannot be directly applied to the engines that use working fluids made up of few particles. In that case, the conventional (or statistical) notion of average quantities such as energy or entropy becomes incomplete. The situation becomes further constrained for engines operating in the quantum regime, where the working fluid is composed of few quantum systems and the effects due to quantum fluctuations cannot be ignored. There have been extensive studies to understand thermodynamics in this regime, see for example \cite{Binder18, Jarzynski97, Crooks99, Campisi11, Alhambra16, Aberg18, Brandao13, Horodecki13, Skrzypczyk14, Brandao15, Lostaglio15a, Bera16, Gour2018, Muller18, Sparaciari17, Bera17, Uzdin18, Khanian20}, and it is revealed that, in general, a quantum system in contact with a thermal bath delivers fluctuating work. Consequently, a quantum engine, where a working fluid sequentially or incoherently interacts with two baths in a cycle in the presence of highly fluctuating input and output energy fluxes, is expected to have fluctuations in both efficiency and power, see for example \cite{Kosloff14, Uzdin15, Klaers17, Rossnagel14, Verley14, Funo15, Ng17, Woods19, Manikandan19,  Esposito10, Scully11, Campisi16, Holubec16a, Brandner17, Denzler20, Abah12, Esposito10a, Guo13, Ma18a, Pietzonka18, Holubec18, Dorfman18, Abiuso20, Brandner20, Miller20, Singh20, Benenti20, Rossnagel16, Saryal2021, Saryal2021a}. In fact, finite size heat engines, in general, deliver fluctuating efficiency \cite{Verley14, Funo15, Manikandan19}. It is also true for power for engines operating with finite-time cycle \cite{Denzler20, Holubec18}. Apart from that, there are power losses due to energy coherence \cite{Brandner17}. Also, there are proposals that smartly exploit energy coherence to increase power, see for example \cite{Scully11, Rossnagel14, Dorfman18}. The overall performance of an engine considering inter-relations between power and efficiency in the presence of quantum fluctuation are studied in \cite{Esposito10a, Guo13, Pietzonka18, Holubec18, Dorfman18, Abiuso20, Brandner20, Miller20, Singh20, Benenti20}. In \cite{Holubec16a}, the authors derive the lower and upper bound on maximum efficiency at a given power for the low dissipation heat engines. This bound generalizes the bound on efficiency at maximum power given by \cite{Esposito10a}. These engines also exhibit universal constraints for efficiency and power \cite{Ma18a}. For general cases, a universal trade-off between efficiency and power is introduced in \cite{Pietzonka18}. A trade-off relation based on geometric arguments is derived in \cite{Brandner20} for any thermodynamically consistent microdynamics. Further studies based on the geometry of work fluctuation and efficiency are made in \cite{Miller20} for microscopic heat engines. 

As in classical engines, it is now commonly believed that yielding maximum power at Carnot efficiency is impossible in a quantum engine. Earlier studies have assumed engines with a working system that interacts with the hot and cold baths at different stages of an engine cycle or with both baths simultaneously but only enabling incoherent heat transfer. Furthermore, the working system is either composed of a statistically large number of particles, or a few particles, allowing a large number of measurements. The role of quantum fluctuations in delimiting efficiency or power or both becomes more prominent for the quantum engines operating in the one-shot finite-size regime, i.e., engines with a finite number of quantum particles constituting the working system and restricted to one-shot measurements or observations. So far, there are no comprehensive studies on that.

Here we introduce quantum heat engines operating in the one-shot finite-size regime and study the power and efficiency of heat-to-work conversion. We show that these engines can simultaneously attain maximum efficiency, i.e., the Carnot efficiency, and maximum power in the one-shot finite-size regime. Therefore, there is no fundamental trade-off between power and efficiency that an engine has to respect in the quantum regime. Our approach is fundamentally different from the earlier ones in the sense that: (i) the engines are fully quantum as they operate in the one-shot finite-size regime and allow genuine entanglement between the baths and working system, (ii) the working system simultaneously interacts with the hot and cold baths via semi-local thermal operations, and (iii) the engines run in a one-step cycle.  The framework relies on the resource theory recently developed to establish the thermodynamic laws in quantum heat engines \cite{MLBera21}. The engines deliver maximum power, along with Carnot efficiency, purely because the engines allow a coherent transfer of heat from hot to cold baths by establishing quantum entanglement between the working system and the baths, thereby attaining maximum quantum speed for the reversible state transformation in each engine cycle.  Finally, we also introduce a physically realizable quantum heat engine based on a quantum-optical system.  

The article is organized as follows. In Section~\ref{sec:EngOp}, we briefly describe the thermodynamical transformation with which a quantum heat engine operates in a one-step cycle. For this, we adhere to the resource theory quantum heat engines, recently developed in~\cite{MLBera21}. Then we present our main results in Section~\ref{sec:MaxPMaxE} and~\ref{sec:QObasedHE}. In Section~\ref{sec:MaxPMaxE}, we discuss how quantum heat engines, equipped with the allowed (or resource-free) thermodynamical operations, can deliver maximum power with Carnot efficiency. A physically realizable model of such an engine is outlined in Section~\ref{sec:QObasedHE} based on quantum optical systems. Finally, we conclude with a discussion in Section~\ref{sec:Disc}.  


\section{Engine operating in one-step cycle \label{sec:EngOp}}
We consider an engine composed of two baths $B_1$ and $B_2$ with corresponding Hamiltonians $H_{B_1}$ and $H_{B_{2}}$ and inverse temperatures $\beta_1$ and $\beta_2$ respectively; a bipartite (working) system $S_{12}$ with non-interacting subsystems $S_1$ and $S_2$ and described by the Hamiltonian $H_{S_{12}}=H_{S_1}+H_{S_2}$; a bipartite battery $S_{W_{S_{12}}}$ with non-interacting subsystems $S_{W_1}$ and $S_{W_2}$ with the Hamiltonian $H_{S_{W_{12}}}=H_{S_{W_{1}}}+H_{S_{W_{2}}}$. Here the battery plays the role of a piston in a traditional engine that takes away work converted from heat. Throughout this work, we assume $\beta_1 < \beta_2$. All systems under consideration have Hamiltonians bounded from below, with the lowest energy equal to zero. The baths are considerably large compared to the systems, and the degeneracy in their microcanonical ensembles scales exponentially with the change in energy. That means the energies of the working systems and the battery are tiny compared to the baths, while the latter have the highest energies close to infinity. The properties of large baths are outlined in Appendix. 

The engine lets the baths interact with the working system and the batteries via a global unitary evolution ($U$) where the composite $S_1S_{W_1}$ semi-locally interacts with $B_1$ and $S_2S_{W_2}$ with $B_2$. As a result, a semi-local thermal operation (SLTO) is implemented on the system-battery composite  $S_{12}S_{W_{12}}$ given by~\cite{MLBera21}
\begin{align} \label{eqmt:slto-Stinespring}
		\Lambda_{S_{12}S_{W_{12}}}& \left(\rho_{S_{12}} \otimes \rho_{S_{W_{12}}}\right) \nonumber \\
		& =\tr_{B_1 B_2} \left[ U (\gamma_{B_1} \otimes \gamma_{B_2} \otimes \rho_{S_{12}} \otimes \rho_{S_{W_{12}}}) U^\dag \right],
	\end{align} 
where the global unitary $U$ satisfies
	\begin{align}
		&\left[U, \ H_{B_1} + H_{S_1} + H_{S_{W_1}} +  H_{B_2} + H_{S_2} + H_{S_{W_2}}  \right] =0, \label{eqmt:slto-commutation} \\
		&\left[U, \ \beta_1 \ (H_{B_1} + H_{S_1}+ H_{S_{W_1}}) + \beta_2 \ (H_{B_2} + H_{S_2}+ H_{S_{W_2}})  \right] =0 \label{eqmt:slto-commutationTemp}.
	\end{align}
Here the baths are in the equilibrium states denoted by $\gamma_{B_x}=\frac{e^{-\beta_xH_{B_x}}}{\tr [e^{-\beta_xH_{B_x}}]}$ for $x=1,2$, and $\rho_{S_{12}}$ is any state of $S_{12}$. The $\rho_{S_{W_{12}}}$ is the state of the battery $S_{W_{12}}$, where the subsystems $S_{W_1}$ and $S_{W_2}$ always remain in their energy eigenstates and store or supply energy in the form of work. The commutation relation \eqref{eqmt:slto-commutation} guarantees strict conservation of total energy of baths-system-battery composite. Note, this ensures conservation of all moments of energy and not just the average energy. The relation \eqref{eqmt:slto-commutation}, in turn, represents the quantum version of first law for engines. 

The relation~\eqref{eqmt:slto-commutationTemp} ensures strict weighted energy conservation. We may place two arguments to justify why quantum heat engines must satisfy this weighted energy conservation. First, this relation is essential for the construction of resource theory \cite{MLBera21}, ensuring that $\Lambda_{S_{12}} \left(\gamma_{S_1} \otimes \gamma_{S_2}\right)=\gamma_{S_1} \otimes \gamma_{S_2}$ where $\gamma_{S_x}=\frac{e^{-\beta_xH_{B_x}}}{\tr [e^{-\beta_xH_{S_x}}]}$ for $x=1, 2$. Here we ignore the batteries as they only store or supply without directly interacting with the baths. This implies that if the subsystems $S_1$ and $S_2$ are in (local) thermal equilibrium with the baths $B_1$ and $B_2$ respectively (also called semi-Gibbs state), then the SLTOs cannot transform them out of the equilibrium. Note, this is a direct consequence of the relation~\eqref{eqmt:slto-commutationTemp}. Thus, these semi-Gibbs states are resource-free states from the thermodynamic point of view. This also physically makes sense. Because if the subsystems are thermalized to the baths (as mentioned above), nothing interesting can happen in terms of energy (heat or work) exchange as constrained by zeroth law. Thus, the local system will remain unaltered. This is also true the traditional heat engines operating in a four-step Carnot cycle. The second argument can be placed as follows. Here we implicitly assume baths are considerably larger than the working system, and the working system returns to its initial state at the end of each cycle. Let say, if the hot bath $B_1$ releases $\Delta S$ amount of entropy then an associated energy needs to flow out from the hot bath is given by $\Delta E_1$, where $ \Delta S = \beta_1 \Delta E_1 $. Since baths and systems together form an isolated composite, the same amount of entropy $\Delta S$ has to be absorbed by the cold bath because the entire process is strictly entropy conserving. This will increase the energy by an amount $\Delta E_2$, where $\Delta S=\beta_2 \Delta E_2$. The entropy conservation implies $\beta_1 \Delta E_1 + \beta_2 \Delta E_2=0$, which is exactly the condition demanded by the commutation relation~\eqref{eqmt:slto-commutationTemp}. Note this argument above is based on the notion of average entropy (Gibbs or von Neumann entropy), and it does not precisely capture the notion of entropy in the one-shot finite-size regime. To understand how the relation~\eqref{eqmt:slto-commutationTemp} ensures strict conservation of total entropy in the one-shot regime, we need to turn to the microscopic picture based on how degeneracy in the micro-canonical ensembles changes due to an energy transfer from the hot to the cold composites. We discuss that in the Appendix. The SLTOs satisfy several interesting properties and can be found in \cite{MLBera21}. 

\begin{figure}
	\includegraphics[width=0.8\columnwidth]{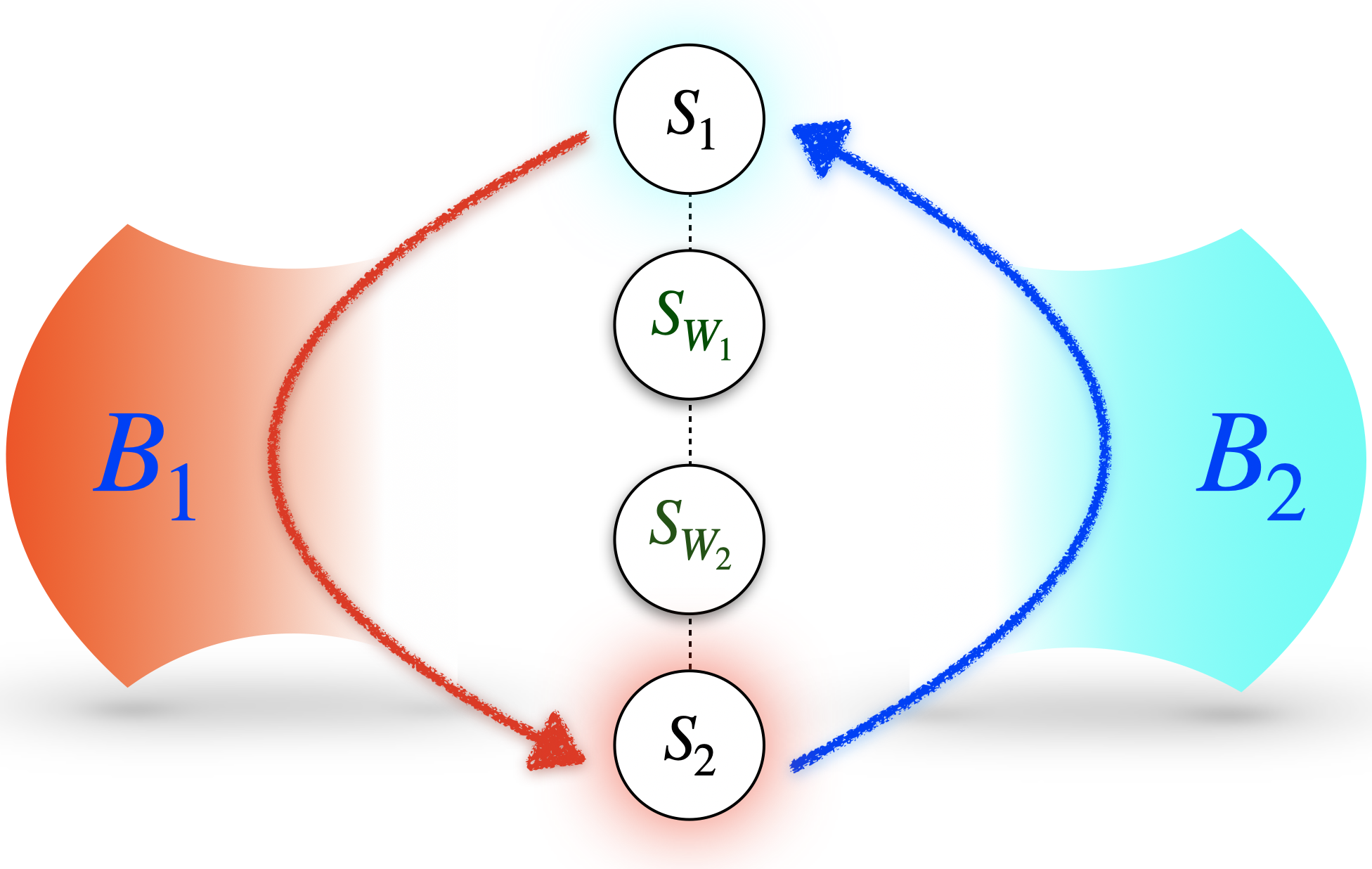}
	\caption{{\bf One-step engine cycle.} An engine consists of two baths $B_1$ and $B_2$ at inverse temperatures $\beta_1$ and $\beta_2$ ($\beta_1 < \beta_2$), a working system $S_{12} \equiv S_1S_2$ and a battery $S_{W_{12}}\equiv S_{W_1}S_{W_2}$. In each one-step engine cycle, the composite $S_1S_{W_1}-S_{W_2}S_2$ semi-locally interact with the baths $B_1-B_2$ and undergoes a transformation so that the working sub-systems $S_1$ and $S_2$ swaps their states along with Hamiltonians (as indicated by the arrows) and the battery sub-systems update their states. As a result, there is an overall flow of heat from $B_1$ to $B_2$ and, in this process, part of that heat is converted into work and stored in $S_{W_1}S_{W_2}$. At the end of each cycle, the state of $S_1S_2$ becomes identical to its initial state upto a swap operation and is, again, reused in the next cycles. See text for more details.
		\label{fig:OneStepCycle}}
\end{figure}

In an engine operating in cycles, the system $S_{12}$ mediates the heat transfer from $B_1$ to $B_2$, while a part of that is converted into work and stored in the battery $S_{W_{12}}$. At the end of each cycle, the $S_{12}$ should recover its initial state so it can be reused for the next cycle. But the battery gets excited to a higher energy eigenstate to store work. Interestingly, the engine executes this transformation in a one-step cycle (see Figure~\ref{fig:OneStepCycle}) by implementing semi-local thermal operations on $S_{12}S_{W_{12}}$, as  
\begin{align}
	\left(\rho_{S_{12}} \otimes \rho_{S_{W_{12}}}^i, \ H_{S_{12}} + H_{S_{W_{12}}} \right) \rightarrow \left( \sigma_{S_{12}}^\prime \otimes \rho_{S_{W_{12}}}^f, \ H_{S_{12}}^\prime + H_{S_{W_{12}}}  \right). \nonumber 
\end{align}
Consequently, the state of the working system transforms as $\rho_{S_{12}} \to \sigma^\prime_{S_{12}}$ and at the same time the Hamiltonian is modified as $H_{S_{12}}=H_{S_1} + H_{S_2} \to H_{S_{12}}^\prime=H_{S_1}^\prime + H_{S_2}^\prime$. Further, it satisfies the cyclicity conditions  $\sigma_{S_{12}}^\prime=U_{S_1 \leftrightarrow S_2} \left( \rho_{S_{12}} \right), \ \ H_{S_1}^\prime=H_{S_2}, \ \ \mbox{and} \ \ H_{S_2}^\prime=H_{S_1}$, where the unitary $U_{S_1 \leftrightarrow S_2}$ swaps the states of the subsystems $S_1$ and $S_2$.  The battery undergoes the transformation  $\rho_{S_{W_{12}}}^i \to \rho_{S_{W_{12}}}^f$ without updating its Hamiltonian.

To understand how the above transformation executes the (four-step) Carnot cycle in one-step, let us focus on the transformation happening in the system, that is $(\rho_{S_{12}}, H_{S_{12}}) \to (\sigma_{S_{12}}^\prime, \ H_{S_{12}}^\prime)$. For this purpose, we ignore the battery as it only changes states without updating its Hamiltonians and thereby stores or releases work. Consider, $\rho_{S_{12}}=\rho \otimes \sigma$ and $H_{S_{12}}=H + H^\prime$, where $H$ and $H^\prime$ are the Hamiltonians of the subsystems $S_1$ and $S_2$ respectively. Then the (one-step) engine operation leads to 
\begin{align}
(\rho \otimes \sigma, \ H + H^\prime ) \to (\sigma \otimes \rho, \ H^\prime + H).	
\end{align}
This involves two simultaneous sub-transformations. One is $(\rho, \ H) \to (\sigma, \ H^\prime)$ via a semi-local interaction with $B_1$, which can be understood as the combination of an isothermal $(\rho, \ H) \to (\sigma, \ H)$ and then an adiabatic $(\sigma, \ H) \to (\sigma, \ H^\prime)$ transformations.  The other sub-transformation $(\sigma, \ H^\prime) \to (\rho, \ H)$ takes place in semi-local interaction with the bath $B_2$, which again can be understood as the combination of  an isothermal $(\sigma, \ H^\prime) \to (\rho, \ H^\prime)$ and then an adiabatic $(\rho, \ H^\prime) \to (\rho, \ H)$ transformations. Clearly, this mimics the situation of a Carnot engine where one working system initially in $(\rho, H)$ undergoes two isothermal (in interaction with two different baths) and two adiabatic transformations, but in one-step (see \cite{MLBera21} for more details).   

\section{Maximum power with Carnot efficiency \label{sec:MaxPMaxE}}
The engines equipped with SLTOs can yield better performance than the traditional heat engines. Not only can the engines execute the Carnot cycle in one step, but they are also superior to conventional heat engines in efficiency and power. Most importantly, these engines can deliver maximum power with Carnot efficiency. Note, to attain maximum power and efficiency simultaneously, the engine has to undergo the fastest possible thermodynamically reversible transformation in each cycle, which we are going to demonstrate below.

Without loss of generality, we consider the working subsystems $S_1$ and $S_2$ are to be qubits with the Hamiltonians $H_{S_1}=a\ketbra{1}{1}_{S_{1}}$ and $H_{S_2}=a\ketbra{1}{1}_{S_{2}}$ respectively having identical energy spacing. We also assume, without loss of generality, that the battery subsystems $S_{W_1}$ and $S_{W_2}$ are qubits with the Hamiltonians $H_{S_{W_{1}}}=E_{W_1} \ketbra{1}{1}_{S_{W_1}}$ and $H_{S_{W_{2}}}=E_{W_2} \ketbra{1}{1}_{S_{W_2}}$ respectively. The maximum heat-to-work conversion efficiency per (one-step) cycle is attained by implementing a thermodynamically reversible state transformation in $S_1S_2S_{W_1}S_{W_{2}}$ composite
\begin{align}
	\ket{0, 1, 0, 0}_{S_1S_2S_{W_1}S_{W_2}} \to \ket{1, 0, 1, 1}_{S_1S_2S_{W_1}S_{W_2}}, \label{eq:SWtrans}
\end{align}
using a semi-local thermal operation \cite{MLBera21}, where the subsystems $S_1$ and $S_2$ swap their states without changing the Hamiltonians, and the batteries $S_{W_{1}}$ and $S_{W_{2}}$ get excited. Here we denote $\ket{i, j, k,l}_{S_1S_2S_{W_1}S_{W_2}}=\ket{i}_{S_1} \otimes \ket{j}_{S_2}\otimes \ket{k}_{S_{W_1}}\otimes \ket{l}_{S_{W_2}}$. 

\begin{figure}
	\includegraphics[width=0.7\columnwidth]{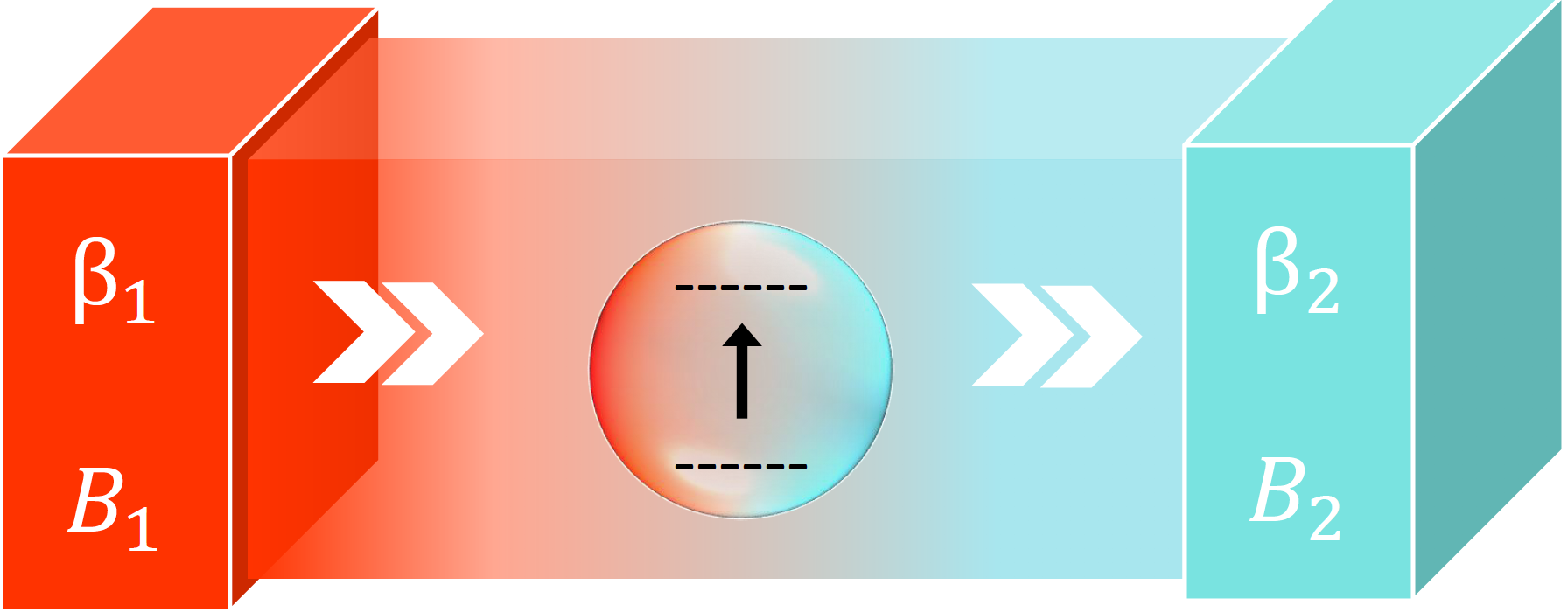}
	\caption{{\bf A compact engine.} An engine consists of two baths $B_1$ and $B_2$ at inverse temperatures $\beta_1$ and $\beta_2$ respectively, and a working system $S$. The system $S$ simultaneously interacts with both the baths via a semi-local thermal operation. The engine operates in a one-step cycle by exciting the system $S$ from lower energy to a higher energy eigenstate.  See text for more details.
		\label{fig:QHEcopy}}
\end{figure}

For simplicity, we may consider the working system and the battery to be the parts of single system $S \equiv S_1S_2S_{W_1}S_{W_2}$ with the Hamiltonian $H_{S}=a_0\ketbra{0}{0}_S + a_1\ketbra{1}{1}_S$, where $a_0=a$ and $a_1=a+E_{W_1} + E_{W_2}$ with the corresponding energy eigenstates $\ket{0}_{S}=\ket{0, 1, 0, 0}_{S_1S_2S_{W_1}S_{W_2}}$ and $\ket{1}_{S}=\ket{1, 0, 1, 1}_{S_1S_2S_{W_1}S_{W_2}}$. Then, the engine becomes compact and has three constituents; hot and cold baths ($B_1$ and $B_2$) and a two-level system ($S$). The engine cycle starts with the initial state $\ket{0}_S$ and ends with the final state $\ket{1}_S$ of $S$ (as shown in Figure \ref{fig:QHEcopy}). The corresponding global transformation, leading to this one-step cycle, is 
\begin{align}
	\gamma_{B_1} \otimes \gamma_{B_2} \otimes \ketbra{0}{0}_S \xrightarrow{U} \tau_{B_1 B_2} \otimes \ketbra{1}{1}_S, \label{eq:OverAllST}
\end{align}  
where $\tau_{B_1 B_2}$ is final state of the baths. The unitary $U$ strictly conserves energy of $B_1B_2S$ composite and weighted-energy of $B_1B_2$ composite, i.e.,
\begin{align}
	&\left[U, \ H_{B_1} +  H_{B_2} + H_{S}  \right] =0, \label{eqmt:slto-commutation-bbs} \\
	&\left[U, \ \beta_1 H_{B_1} + \beta_2 H_{B_2}  \right] =0. \label{eqmt:slto-commutationTemp-bbs}
\end{align}
The commutation relation~\eqref{eqmt:slto-commutation-bbs} ensures the strict conservation of total energy. The relation~\eqref{eqmt:slto-commutationTemp-bbs} ensuring strict conservation of entropy is a special case of the general condition~\eqref{eqmt:slto-commutationTemp} where the transformation is cyclic. Here the system $S$ remains in the energy eigenstates before and after the cycle without changing its energy and entropy. Because of that, the entropy conservation is guaranteed by the strict weighted-energy conservation of the baths only. See Appendix for more details. Note, the semi-local nature of the evolution is clearly understood here as the system $S$ is simultaneously interacting with both the baths via the unitary $U$.

Now we show that the unitary $U$, that satisfies relations \eqref{eqmt:slto-commutation-bbs} and \eqref{eqmt:slto-commutationTemp-bbs}, leading to the transformation~\eqref{eq:OverAllST} indeed attains Carnot efficiency. Then we consider constructing a driving Hamiltonian, corresponding to the unitary $U$, that delivers maximum power with Carnot efficiency.

Since the initial (and final) state of $S$ is an energy eigenstate, the global initial state of $B_1B_2S$ can be expressed in the block-diagonal form with respect to total energies, as
\begin{align}
	\gamma_{B_1} \otimes \gamma_{B_2} \otimes \ketbra{0}{0}_S =\bigoplus_{E_{B_{12}}+a_0}[\gamma_{B_1} \otimes \gamma_{B_2}]_{E_{B_{12}}} \otimes \ketbra{0}{0}_S,
\end{align} 
with  $E_{B_{12}}=E_{B_1}+E_{B_2}$, where $E_{B_1}$ and $E_{B_2}$ are the energies corresponding to the baths $B_1$ and $B_2$ respectively, and
\begin{align}
	[\gamma_{B_1} \otimes \gamma_{B_2}]_{E_{B_{12}}}= p(E_{B_{12}})  \sum_{i=1}^{d_1({E_{B_1}})d_2({E_{B_2}})}\ketbra{E_{B_{12}}(i)}{E_{B_{12}}(i)}, \nonumber
\end{align} 
where $d_1({E_{B_1}})$ and $d_2({E_{B_2}})$ represent the degeneracies corresponding to the bath energies $E_{B_1}$ and $E_{B_2}$ respectively  and $p(E_{B_{12}})=e^{-\beta_1 E_{B_1} -\beta_2 E_{B_2}}/Z_{B_1}Z_{B_2}$. A strictly total energy conserving unitary also takes a block-diagonal form, $U=\bigoplus_{E_{B_{12}+a_0}}U_{E_{B_{12}+a_0}}$, where the unitary $U_{E_{B_{12}+a_0}}$ operates only on the block with the total energy $E_{B_{12}}+a_0$ and implements a transformation
\begin{align}
[\gamma_{B_1} \otimes \gamma_{B_2}]_{E_{B_{12}}} \otimes \ketbra{0}{0}_S \to [\tau_{B_1B_2}]_{E^\prime_{B_{12}}} \otimes \ketbra{1}{1}_S,	\label{eq:BlockTrans}
\end{align}
where $E_{B_{12}}^\prime = E_{B_1}^\prime + E_{B_2}^\prime$. Note, $E_{B_{12}} + a_0 = E_{B_{12}}^\prime + a_1$ as required by the total energy conservation. The strict conservation of total weighted-energy of the baths ensures
\begin{align}
\beta_1 (E_{B_1}^\prime-E_{B_1}) + \beta_2 (E_{B_2}^\prime - E_{B_2})=\beta_1 Q_1 + \beta_2 Q_2=0,	\label{eq:ClausiusEqu}
\end{align}
where $Q_1$ and $Q_2$ are the heat flow out of the baths $B_1$ and $B_2$ respectively. The~\eqref{eq:ClausiusEqu} is nothing but the Clausius equality. This in turn ensures the thermodynamic reversibility of the state transformation. Note, this condition also implies $d_1(E_{B_1}) d_2(E_{B_2})=d_1(E_{B_1}^\prime) d_2(E_{B_2}^\prime)$, where $d_1(E_{B_1}^\prime)$ and $d_2(E_{B_2}^\prime)$ are degeneracies in the energies corresponding to $E_{B_1}^\prime$ and $E_{B_2}^\prime$ of the baths $B_1$ and $B_2$ respectively (see Appendix for more details). It is an essential requirement for a unitary transformation where the rank and the spectra of the (un-normalized) state of each total energy block remain unchanged. 

Similar transformations, as in Eq.~\eqref{eq:BlockTrans}, also take place in all other total energy blocks due to the evolution by the unitary $U$. As a result, the desired state transformation, given in Eq.~\eqref{eq:OverAllST}, is achieved and thereby completes the one-step engine cycle. The extracted work per cycle is given by $W_{ext}=a_1-a_0=Q_1 + Q_2$ as a consequence of strict conservation of total energy.  Hence, the heat-to-work conversion efficiency becomes maximum in the one-shot finite-size regime, given by
\begin{align}
\eta=\frac{W_{ext}}{Q_1}=1-\frac{\beta_1}{\beta_2},
\end{align}
which is the Carnot efficiency, as expected for any reversible engine cycle. 

Let us demonstrate how the global unitary $U$ can be implemented using an interaction Hamiltonian
\begin{align}
	H_{in}= \hbar g \bigoplus_{E_{B_{12}}+a_0} \sum_{i=1}^{d_1(E_{B_1}) d_2(E_{B_2})}  \ketbra{E_{B_{12}}^\prime(i), 1}{E_{B_{12}}(i), 0}_{B_1B_2S} + h.c., \nonumber
\end{align}
where, again, $\ket{E_{B_{12}}(i)} \equiv \ket{E_{B_1}(i), E_{B_2}(i)}$ and $\ket{E_{B_{12}}^\prime(i)}\equiv \ket{E_{B_1}^\prime(i), E_{B_2}^\prime(i)}$, and $g$ is a constant. The global unitary is then $U(t)=e^{-i  t H_{in}/\hbar}$ for any time $t$. Under this unitary, an initial state $\ket{E_{B_1}(i), E_{B_2}(i), 0}_{B_1B_2S}$ in the total energy block evolves to  $\ket{\psi(t)}= U(t) \ket{E_{B_1}(i), E_{B_2}(i), 0}_{B_1B_2S}$ at time $t$, where
\begin{align}
	\ket{\psi(t)} = \cos(gt) \ \ket{&E_{B_1}(i), \ E_{B_2}(i), \ 0}_{B_1B_2S} \nonumber \\
	&-i \sin (gt) \ \ket{E_{B_1}^\prime(i), \ E_{B_2}^\prime(i), \ 1}_{B_1B_2S}, \label{eq:EntangledState}	
\end{align}
which is a genuinely entangled state of $B_1$, $B_2$, and $S$ for $gt \neq z \pi/2 $ with $z \in \mathbb{Z}$. The desired final state is attained at time $\tau=\pi/(2g)$, where all the constitutes become uncorrelated from each other. It is important to highlight that the above engine evolution enables a coherent heat transfer from $B_1$ to $B_2$, which happens due to entanglement in the intermediate time and is fundamentally different from conventional engines. Similar evolution takes place in every total energy block, and the overall transformation~\eqref{eq:OverAllST} is attained at time $\tau$. With this, the engine extracts $W_{ext}$ work in $\tau$ time. Thus, the power delivered by the engine, i.e., work extraction per unit time, is
\begin{align}
	P=\frac{W_{ext}}{\tau}=\frac{2gW_{ext}}{\pi}. \label{eq:MaxPower}
\end{align}  

\begin{figure}
	\includegraphics[width=0.80\columnwidth]{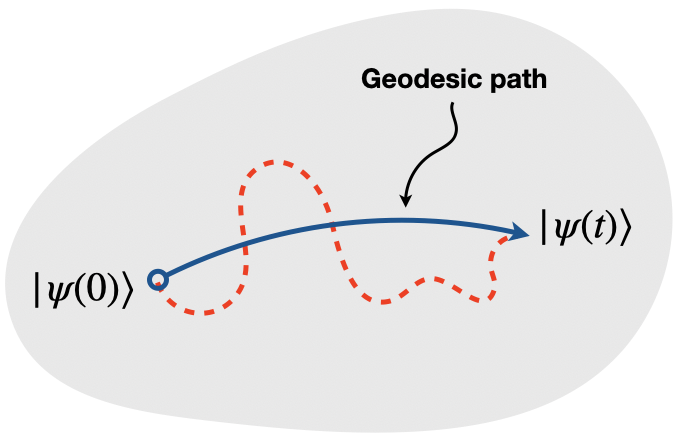}
	\caption{{\bf Geodesic trajectory of evolution.} Evolution of states on a quantum state space (the gray area). An initial state $\ket{\psi(0)}$ at time $t=0$ is evolved to $\ket{\psi(t)}$ at time $t=t$. There are infinitely many paths connecting the states. One with the shortest path is called the geodesic path (solid (blue) line). Any other path (dotted (red) line) is longer than the geodesic one.  
		\label{fig:Geodesic}}
\end{figure}

Contrary to the traditional understanding, the reversible (one-step) engine cycle via semi-local thermal operation requires finite time. Not only that, as we argue below, the interaction Hamiltonian $H_{in}$ drives the evolution with the maximum attainable speed to result in the shortest possible transformation time. The speed of evolution is defined by the distance traversed by a system per unit time in its quantum state space \cite{Anandan90}. For pure states, the distance is measured using Fubini-Study metric, given by $s=\frac{1}{2}(1-|\braket{\psi | \phi}|^2)$ for any two states $\ket{\psi}$ and $\ket{\phi}$. The speed of evolution of the state $\ket{\psi(t)}$ is 
\begin{align}
	v=\frac{ds}{dt}=\frac{\Delta H_{in}}{\hbar}=g,
\end{align}
where $ds=\frac{1}{2}(1-|\braket{\psi(t) | \psi(t + dt)}|^2)$ and energy uncertainty $\Delta H_{in} = \sqrt{\braket{\psi(t)|H_{in}^2|\psi(t)}  - \braket{\psi(t)|H_{in}|\psi(t)}^2}$. Note, the speed of evolution $v$ is same for every total energy block and, hence, for the overall transformation. The uncertainty $\Delta H_{in}$ is independent of time and has the maximum possible value, equals to $\hbar g$  for any driving Hamiltonian bounded by the operator norm $\hbar g$. Furthermore, the interaction Hamiltonian $H_{in}$ drives the evolution following a geodesic trajectory \cite{Anandan90} connecting the initial and the final state which represents the shortest path (see Figure \ref{fig:Geodesic}). The evolution following shortest path with maximum speed results in the minimum required time to complete the transformation in the one-step engine cycle. As a consequence, the power $P$ in Eq.~\eqref{eq:MaxPower} is the maximum possible one. Note, quantum effects such as entanglement are believed to degrade the performance of engines. But, on the contrary, here we find that the engines operating with semi-local thermal operations can exploit entanglement to deliver maximum power with Carnot efficiency.     

\section{A quantum optics based heat engine \label{sec:QObasedHE}}
\begin{figure}
	\includegraphics[width=0.95\columnwidth]{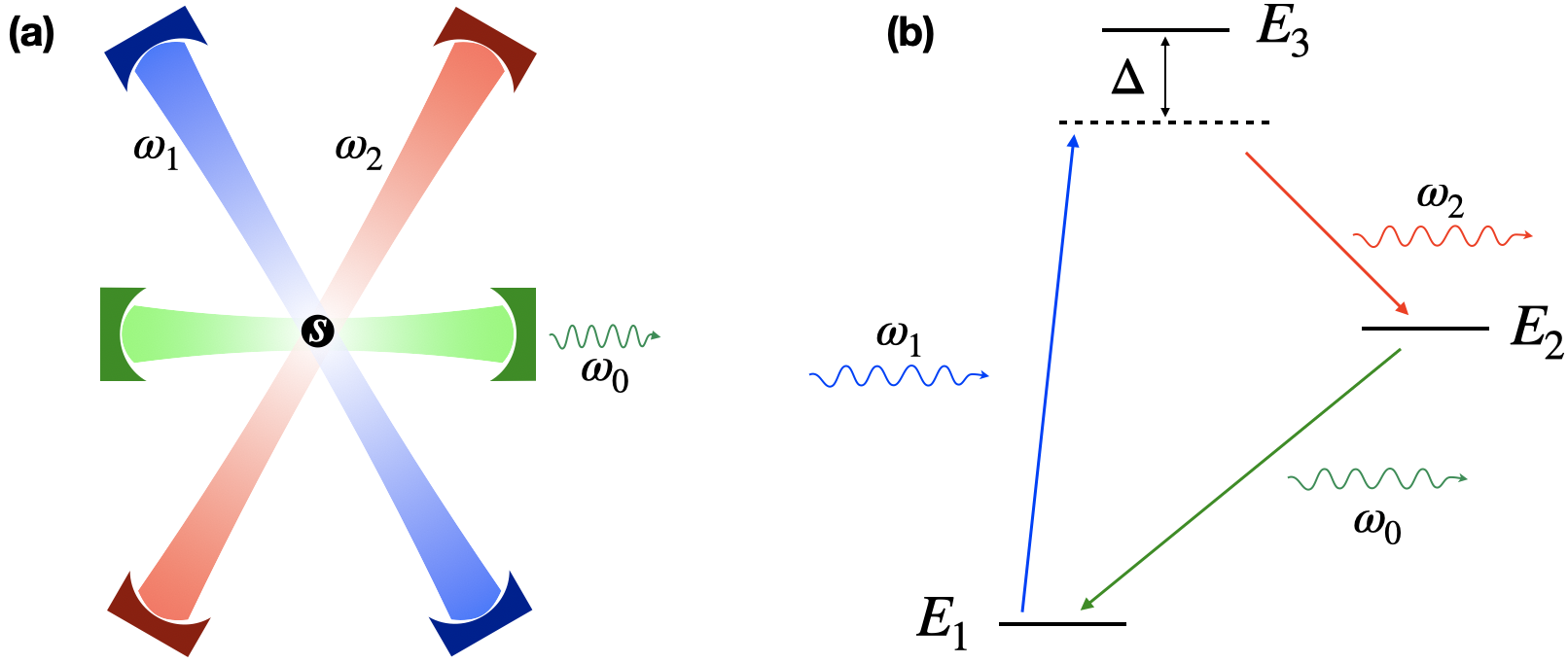}
	\caption{{\bf An optical-cavity based quantum heat engine.} (a) A three-level quantum system $S$ (i.e., working system) is placed in three overlapping optical cavities. Thermal cavities with frequencies $\omega_1$ and $\omega_2$ at inverse temperatures $\beta_1$ and $\beta_2$ ($\beta_1 < \beta_2$) represent the baths $B_1$ and $B_2$ respectively. The cavity with frequency $\omega_0$ is in resonance with the transition between the ground state and the first excited state of the system. (b) The three-level system $S$ simultaneously interacts with baths $B_1$ and $B_2$ via two-mode amplitude-dependent coupling. In each engine cycle, it absorbs a photon with energy $\hbar \omega_1$ from $B_1$ and emits a photon with $\hbar \omega_2$ energy to the bath $B_2$ and excites itself from energy $E_1$ to $E_2$. The system then emits a photon with energy $\hbar \omega_0$ via stimulated emission in a cavity (green) in resonance with the transition. See text for more details.
		\label{fig:Lambda_system}}
\end{figure}

Here we discuss a physically realizable quantum heat engine transferring maximum power with Carnot efficiency following the theoretical framework presented above. We propose an engine composed of two thermal cavities and a three-level working system (see Figure~\ref{fig:Lambda_system}). The bath $B_1$ is a single-mode optical cavity with a Hamiltonian $H_{B_1}=\hbar \omega_1 a_1^\dag a_1=\sum_{n} n \omega_1 \ketbra{n}{n}_{B_1}$, at inverse temperature $\beta_1$. Here $a_1^\dag$ and $a_1$ are the creation and annihilation operators of the mode in $B_1$ respectively, $\omega_1$ represents the mode frequency, and $n$ and $\ket{n}$ are the number of excitation and the corresponding number state. Similarly, the bath $B_2$ at inverse temperature $\beta_2$ is another optical cavity with a Hamiltonian $H_{B_2}=\hbar \omega_2 a_2^\dag a_2=\sum_{m} m \omega_2 \ketbra{m}{m}_{B_2}$. The system $S$ is a three-level atom (in $\Lambda$-configuration) with the Hamiltonian $H_S=\sum_{i=1}^3 E_i \ketbra{i}{i}_S$ with $E_1=0$.  The overall Hamiltonian of the baths and the system composite is then $H_0=H_{B_1} + H_{B_2}+H_S$.

A semi-local thermal operation, leading to a one-step cycle, is implemented by introducing an intensity-dependent coupling between the bath modes and the atom by the interaction Hamiltonian
\begin{align}
	H_I=  f_1(N_1) + & f_2(N_2) +  \hbar g_1 \theta_1(N_1) (a_1 \sigma_{31} + h.c.)  \nonumber \\ 
	&+ \hbar g_2 \theta_2(N_2) (a_2 \sigma_{32} + h.c.), \label{eq:Hi}
\end{align}
with the number operator $N_k=a_k^\dag a_k$ corresponds to the bath $B_k$ for $k=1,2$, and $\sigma_{ij}=\ketbra{i}{j}_S$ ($i \neq j$) is the transition operator from $\ket{j}_S$ to $\ket{i}_S$ for $i, j=1, 2, 3$. The $f_1(N_1)$ and $f_2(N_2)$ are some intensity-dependent potentials in the cavity fields. The state $\ket{3}_S$ is coupled with $\ket{1}_S$ and $\ket{2}_S$ via intensity-dependent dipole-couplings $g_1\theta(N_1)$ and $g_2 \theta(N_2)$ respectively, where $\theta(N_1)$ and $\theta(N_2)$ are some functions of the number operators, and $g_1$ and $g_2$ are some constants. There is no direct coupling between $\ket{1}_S$ and $\ket{2}_S$. The technical details of how the interaction Hamiltonian \eqref{eq:Hi} may be realized are described in Appendix.

With the choice of (identical) detuning $\Delta=(E_3 - E_k)/\hbar - \omega_k$ and the couplings
\begin{align}
\frac{g_k^2}{\Delta} \theta_k^2(N_k)=f_k(N_k)=\frac{g_k^2}{\Delta} N_k^{-1},	\label{eq:ftheta}
\end{align}
for $k=1, \ 2$. The above formula should be fulfilled possibly exactly for large values of $N_k$, especially if we work at relatively high temperatures. It has to be regularized, obviously, for $N_k=0$ to avoid the singularity. Nevertheless, with these choices, the three-level problem can be exactly reduced to a two-level problem irrespective of whether the detuning $\Delta$ is small or large, similar to what is shown in \cite{Gerry90, Wu96, Greentree2013}.  Then, the corresponding two-level Hamiltonian becomes $H_S^\prime = \frac{1}{2}\hbar \omega_0 (\ketbra{2}{2} - \ketbra{1}{1})$, and the interaction Hamiltonian, after rotating-wave approximation, transforms to 
\begin{align}
	H_I^\prime=\hbar g (A_1A_2^\dag \sigma_{12} + A_1^\dag A_2 \sigma_{21}),
\end{align} 
where $A_k=a_k N_k^{-1/2}$ and $g=g_1g_2/\Delta$ is a constant.  Here $\omega_0 \approx \omega_1 - \omega_2$, i.e., the pump mode with $\omega_1$ and Stokes mode with $\omega_2$ are in two-mode resonance with the states $\ket{1}_S$ and $\ket{2}_S$. The $\omega_1$ and $\omega_2$ are chosen so that $\beta_1 \omega_1=\beta_2 \omega_2$. Then, the unitary $U(t)=\exp[-i t H_I^\prime/\hbar]$ generated by $H_I^\prime$ strictly conserves total energy of baths and system, and total weighted-energy of the baths alone, as $[U(t), \ H_S^\prime + H_{B_1} + H_{B_2}]=0$ and  $[U(t), \ \beta_1 H_{B_1} + \beta_2 H_{B_2}]=0$ respectively for all time $t$.  

The initial state of the composite $B_1B_2S$ can now be expressed in blocks classified by $(n,m)$, as
\begin{align}
	\gamma_{B_1} \otimes \gamma_{B_2} \otimes \ketbra{1}{1}_S=\bigoplus_{n,m} p_{nm} \  \ketbra{n,m,1}{n,m,1}_{B_1B_2S}, \nonumber
\end{align}
with $p_{nm}=\exp[-\beta_1 nE_{B_1} -\beta_2 mE_{B_2}]/Z_{B_1}Z_{B_2}$, where $Z_{B_1}$ and $Z_{B_2}$ are the partition functions of the baths $B_1$ and $B_2$ respectively. Due to the constraints on strict total energy conservation, the $U(t)$ operates on each block $(n,m)$ independently. For a block $(n,m)$, the initial state $\ket{n, m, 1 }_{B_1B_2S}$ evolves to $\ket{\phi(t)}=U(t) \ket{n, m, 1 }_{B_1B_2S}$ at some time $t$, and it is given by
\begin{align}
	\ket{\phi(t)}=\cos (g t) \ & \ket{n, m, 1 }_{B_1B_2S}  \nonumber \\ 
	&- i \sin (g t) \ \ket{n-1, m+1, 2}_{B_1B_2S},
\end{align}
which is an entangle state. This is true for all blocks $(n,m)$, except the blocks $(0,m)$. Note, the time taken to evolve the initial state to the desired final states are same for all blocks except $(0, \ m)$, and that is $\tau= \pi/(2g)$. As a consequence, the joint initial state of $B_1B_2S$ evolves to $\rho_{B_1B_2S}^f = U(t) (\gamma_{B_1} \otimes \gamma_{B_2} \otimes \ketbra{0}{0}_S) U(t)^\dag$, and at time $\tau= \pi/(2g)$, the final state of the system $S$ becomes, 
\begin{align}
	\rho_S^f=\tr_{B_1B_2} \ \rho_{B_1B_2S}^f=\frac{1}{Z_{B_1}} \ketbra{1}{1}_S + \left(1-\frac{1}{Z_{B_1}} \right) \ketbra{2}{2}_S \to \ketbra{2}{2}_S \nonumber
\end{align}
for $Z_{B_1} \to \infty$, which is true for low inverse temperature $\beta_1$ of bath $B_1$, or $|B_1| \to \infty$. In each cycle, the the system $S$ undergoes the transformation $\ket{1}_S \to \ket{2}_S$ and thereby extracts $\hbar \omega_0$ amount of work with the Carnot efficiency $\eta=1-\beta_1/\beta_2$. The transformation takes place with the maximum quantum speed following a geodesic trajectory and time requires for that is $\tau=\pi/(2g)$. Hence, the cycle delivers maximum power $P=2 g \hbar \omega_0/\pi$.  The work is extracted in the form of photons at $\omega_0$ by placing the atom in a resonant cavity and letting the stimulated emission $\ket{2}_S \to \ket{1}_S$ (see Fig.~\ref{fig:Lambda_system}).

It is worth mentioning that there have been several propositions of quantum heat engines based on optical cavity or bosonic baths earlier, for example in \cite{Scovil59, Ghosh18}, where a quantum system interacts with two bosonic thermal baths at different temperatures. However, in contrast to the engines considered above, these engines only allow incoherent heat transfer from hot to cold baths. They do not guarantee strict conservation of total energy in order to characterize the energetics correctly. Because of that, they cannot deliver maximum power with Carnot efficiency. 

\section{Discussion \label{sec:Disc}}
For finite-time classical engines, it is known that the maximum power at maximum heat-to-work conversion efficiency is impossible \cite{Curzon75}. For quantum engines, where the working systems interacting with the baths are quantum mechanical, the situation is quite different because the quantum uncertainties present in the system further delimit the extractable work in each cycle. For finite-time quantum heat engines considered earlier, there are various trade-off relations between power and efficiency \cite{Pietzonka18, Brandner20}, and both of these quantities cannot be maximized simultaneously. So far, there are two kinds of quantum engines that have been considered in the literature. In the first kind, the working systems are often composed of a large number of quantum particles. In the second kind, the working system comprises few quantum particles but is allowed to be observed or measured repeatedly for an arbitrarily large number of times. However, the working system interacts with the hot and cold baths in different steps in both kinds.  Hence the heat flow from the hot to the cold baths is not continuous; rather, it occurs in different steps or in an incoherent manner. And, possibly because of this feature of the engines, power and efficiency satisfy trade-off relations and cannot be maximized simultaneously.

The quantum engines considered here can deliver maximum power with maximum efficiency and are fundamentally different from conventional ones studied earlier. Firstly, the engine operates in the one-shot finite-size regime, where the working system is genuinely quantum in the sense that it is made up of a small number of quantum particles (i.e., of finite-size) and allows one or few observations or measurements (i.e., one-shot measurement). Secondly, the working system interacts with both hot and cold baths simultaneously via a semi-local thermal operation. These operations are powerful compared to the operations in traditional engines as they can implement a one-step engine cycle and create entanglement between the baths and the working system. Because of that, it enables a coherent flow of heat from hot to cold bath via the working system and results in maximum power with maximum efficiency. With this, our results have demonstrated that there, in principle, does not exist a fundamental trade-off relation between power and efficiency. We have also put forward an experimentally feasible quantum heat engine operating in the one-shot finite-size regime with a three-level atom as a working system and two thermal optical cavities as the baths. We have explicitly introduced an intensity-dependent interaction between the atom and cavities that executes the one-step engine cycle yielding maximum power at Carnot efficiency.    \\

\noindent In summary: 

\begin{itemize}
\item We have introduced quantum heat engines that operate via one-step cycles in the one-shot finite-size regime and enable a coherent heat transfer from hot to cold baths by establishing genuine quantum entanglement between the working system and the baths.   

\item We have demonstrated that there is no fundamental trade-off relation between power and efficiency. 

\item We have shown a general protocol with which a quantum heat engine can deliver maximum power with Carnot efficiency in the one-shot finite-size regime. 

\item We have proposed a physically realizable model of such a quantum heat engine based on an atom-cavity system. 

\item Our work opens up avenues for an improved theoretical understanding of thermodynamics in the quantum regime and new possibilities for quantum-enabled technologies using heat engines. 
\end{itemize}

{\bf Acknowledgments } -- M.L.B., S.J.-F., and M.L. thankfully acknowledge support from ERC AdG NOQIA, Agencia Estatal de Investigaci\'on (``Severo Ochoa'' Center of Excellence CEX2019-000910-S, Plan National FIDEUA PID2019-106901GB-I00/10.13039/501100011033, FPI), Fundaci\'o Privada Cellex, Fundaci\'o Mir-Puig, and from Generalitat de Catalunya (AGAUR Grant No. 2017 SGR 1341, CERCA program, QuantumCAT\_U16-011424, co-funded by ERDF Operational Program of Catalonia 2014-2020), MINECO-EU QUANTERA MAQS (funded by State Research Agency (AEI) PCI2019-111828-2/10.13039/501100011033), EU Horizon 2020 FET-OPEN OPTOLogic (Grant No 899794), and the National Science Centre, Poland-Symfonia Grant No. 2016/20/W/ST4/00314. M.N.B. gratefully acknowledges financial supports from SERB-DST (CRG/2019/002199), Government of India. 

\appendix

\section{Properties of large baths \label{app:BathProp}}
Our formalism exploits some useful properties of the baths that are considerably large compared to the working systems under consideration. That is why a bath always remains in thermal equilibrium at a fixed temperature even after it interacts with a system. A heat engine has two baths and a working body as components. Hence, a bath $B_x$ with Hamiltonian $H_{B_x}$ is expected to remain in the Gibb's state $\gamma_{B_x}=\frac{e^{-\beta_x H_{B_x}}}{\tr [e^{-\beta_x H_{B_x}}]}$ always, with inverse temperature $\beta_x$. Now for two baths $B_1$ and $B_2$, with corresponding Hamiltonians $H_{B_1}$ and $H_{B_2}$ respectively, the combined state can be expressed as
\begin{align}
	\gamma_{B_{12}}=\gamma_{B_1} \otimes \gamma_{B_2}.
\end{align}
In general, the baths and the working body have Hamiltonians bounded from below, having the lowest energy to be zero. But, the baths may have considerably large energy, i.e., $E^{max}_{B_x} \rightarrow \infty$. Although the combined baths are probabilistic mixtures of all energy states, there exists a set of total energies $\mathcal{E}_{B_{12}}$ where the baths can be found with high probability. Mathematically, it means
\begin{align}\label{eq:BathTypicality}
	\tr [P_{\mathcal{E}_{B_{12}}} \gamma_{B_{12}}] \geqslant 1- \delta,
\end{align}
where $P_{\mathcal{E}_{B_{12}}}$ is a projector that spans over the space corresponding to the set $\mathcal{E}_{B_{12}}$ and $\delta > 0$. With this, the properties of the combined baths are listed below (cf. \cite{MLBera21}).

\begin{itemize}
	\item For any energy $E_{B_{12}}= E_{B_{1}} + E_{B_{2}} \in \mathcal{E}_{B_{12}}$  with$E_{B_{1}}$ and $E_{B_{2}}$ are the energies of the baths $B_1$ and $B_2$ respectively, there is a peak around the mean value $\braket{E_{B_{12}}}$, as $E_{B_{12}} \in \left\{ \braket{E_{B_{12}}} - O(\sqrt{E_{B_{12}}}), \ldots , \braket{E_{B_{12}}} + O(\sqrt{E_{B_{12}}})  \right\}$.  
	
	\item For any energy  $E_{B_{12}}= E_{B_{1}} + E_{B_{2}}\in \mathcal{E}_{B_{12}}$, $E_{S_1} \ll E_{B_1}$, and $E_{S_1}^\prime \ll E_{B_1}$, there exists a $E_{B_{12}}^\prime= E_{B_{1}}^\prime + E_{B_{2}}^\prime \in \mathcal{E}_{B_{12}}$ so that $E_{B_{1}} + E_{S_{1}}=E_{B_{1}}^\prime + E_{S_{1}}^\prime$ and $E_{B_{2}} + E_{S_{2}}=E_{B_{2}}^\prime + E_{S_{2}}^\prime$, where $E_{S_2} \ll E_{B_2}$ and $E_{S_2}^\prime \ll E_{B_2}$.
	
	\item For any energy $E_{B_{1}} + E_{B_{2}}\in \mathcal{E}_{B_{12}}$, the degeneracy $g_B(E_{B_{1}} + E_{B_{2}})$ scales exponentially with energy, and it satisfies the relation $g_B(E_{B_{1}}^\prime + E_{B_{2}}^\prime) \approx g_B(E_{B_{1}} + E_{B_{2}}) \ e^{\beta_1 E_{S_{1}}+\beta_2 E_{S_{2}}}$, where  $E_{B_{1}}^\prime=E_{B_1}+E_{S_1}$ and $E_{B_{2}}^\prime=E_{B_2}+E_{S_2}$ with $E_{S_1} \ll E_{B_1}$ and $E_{S_2} \ll E_{B_2}$.
\end{itemize}

\section{Reversible Engine Operation in a One-step Cycle}
Here we reconsider the reversible engine operation, given in the main text (see Eq.~\eqref{eq:SWtrans}), that yields maximum power with Carnot efficiency. We have assumed a bipartite working system $S_{12}$ with the Hamiltonian $H_{S_{12}}=H_{S_1} + H_{S_2}$ where $H_{S_1}=a \ketbra{1}{1}_{S_1}$ and $H_{S_2}=a \ketbra{1}{1}_{S_2}$. We have also assumed a bipartite battery $S_{W_{12}}$ with the Hamiltonian $H_{S_{W_{12}}}=H_{S_{W_1}}+H_{S_{W_2}}$, where  $H_{S_{W_1}}=E_{W_1}\ketbra{1}{1}_{S_{W_1}}$ and $H_{S_{W_2}}=E_{W_2}\ketbra{1}{1}_{S_{W_2}}$. The one-step cycle is executed by implementing a global unitary ($U$) operation on the baths-system-battery composite leading to the transformation
\begin{align}\label{eq:OveralAllTrans}
	\gamma_{B_1} \otimes \gamma_{B_2} \otimes \rho_{S_{12}} \otimes  \rho^i_{S_{W_{12}}} \to \sigma_{B_1B_2} \otimes \sigma_{S_{12}} \otimes \rho^f_{S_{W_{12}}},  
\end{align}
where $\rho_{S_{12}}=\ketbra{0}{0}_{S_{1}} \otimes \ketbra{1}{1}_{S_{2}}$ and  $\sigma_{S_{12}}=\ketbra{1}{1}_{S_{1}} \otimes \ketbra{0}{0}_{S_{2}}$ are the initial and final states of the working system $S_{12}$, and $\rho_{S_{W_{12}}}^i=\ketbra{0}{0}_{S_{W_1}} \otimes \ketbra{0}{0}_{S_{W_2}}$ and $\rho_{S_{W_{12}}}^f=\ketbra{1}{1}_{S_{W_1}} \otimes \ketbra{1}{1}_{S_{W_2}}$ are the initial and final states of the battery $S_{W_{12}}$. Recall, the global unitary $U$ respects strict conservation of total energy and total weighted-energy. Therefore, we can study the transformation in each total energy block separately. Consider a block of total energy $E_1+E_2$, where $E_1=E_{S_1}+E_{B_1}$ is the sum of energies belonging to $S_1$ and $B_1$, and similarly for $E_2=E_{S_2}+E_{B_2}$. In this total energy block, the transformation becomes
\begin{align}\label{eq:E12Block}
	[\gamma_{B_1} \otimes \gamma_{B_2} \otimes \rho_{S_{12}}]_{E_1+E_2} \otimes  \rho^i_{S_{W_{12}}} \to [\sigma_{B_1B_2} \otimes \sigma_{S_{12}}]_{E_1^\prime + E_2^\prime} \otimes \rho^f_{S_{W_{12}}}, 
\end{align}
where $E_1^\prime=E_{S_1}^\prime+E_{B_1}^\prime$ and $E_2^\prime=E_{S_2}^\prime+E_{B_2}^\prime$. The strict conservation of the total weighted-energy and the total energy ensure that 
\begin{align}
	\beta_1 E_1 + \beta_2 E_2 & = \beta_1 (E_1^\prime + E_{W_1}) + \beta_2 (E_2^
	\prime + E_{W_2}), \label{eq:TotWeightEngCons} \\ 
	E_1 + E_2 & =E_1^\prime +  E_2^\prime + E_{W_1} + E_{W_2}, \label{eq:TotEngCons}
\end{align}
where $E_1=E_{B_1}$, $E_2=E_{B_2}+a$, $E_1^\prime=E_{B_1}^\prime + a$, and $E_2^\prime=E_{B_2}^\prime$. Here we have assumed $\beta_1 < \beta_2$. 

Note, similar transformations will follow in the other total weighted-energy blocks with identical initial and final battery states. The reduced transformation on the system $S_{12}$, $\rho_{S_{12}} \rightarrow \sigma_{S_{12}}$ is reversible because all $\alpha$-free-entropies for pure system and battery states considered here are $\alpha$ independent \cite{MLBera21}. As a consequence, free-entropy distances satisfy $S_d(\rho_{S_{12}} \rightarrow \sigma_{S_{12}} ) = S_d(\rho_{S_{12}} \leftarrow \sigma_{S_{12}} )$, where 
\begin{align}\label{eq:SdCarnotMax}
	S_d(\rho_{S_{12}} \to \sigma_{S_{12}})=\beta_2 a - \beta_1 a = \beta_1 E_{W_1} + \beta_2 E_{W_2} > 0.
\end{align}
This relation guarantees that there is strict conservation of weighted-energy of the working system and the battery together. Therefore, conservation of total weighted-energy~\eqref{eq:TotWeightEngCons} is reduced down to the strict conservation of the weighted-energy of the baths only, i.e.
\begin{align}
	\beta_1 (E_{B_1} - E_{B_1}^\prime) + \beta_2 (E_{B_2} - E_{B_2}^\prime) & = \beta_1 Q_1 + \beta_2 Q_2=0, \label{eq:ClausiusSaturate}
\end{align}
where we have identified the heat as the change in energy of the bath $B_1$ given by $Q_1=E_{B_1} - E_{B_1}^\prime$ and similarly $Q_2=E_{B_2} - E_{B_2}^\prime$ for bath $B_2$. This is true for all energy blocks. The Eq.~\eqref{eq:ClausiusSaturate} represents the \emph{Clausius equality} for the cyclic process. The other total energy blocks will result in identical Clausius equality. The net extracted work in each (one-step) engine cycle  is given by 
\begin{align}
	W_{ext}=E_{W_1}+E_{W_2} =Q_1 + Q_2 >0, \label{eq:maxWork} 
\end{align}
Here we have used the strict total energy conservation \eqref{eq:TotEngCons}. It is clear from Eqs.~\eqref{eq:ClausiusSaturate} and~\eqref{eq:maxWork} that the heat-to-work conversion is $\eta_C=\frac{W_{ext}}{Q_1}=1-\frac{\beta_1}{\beta_2}$, which is exactly the Carnot efficiency. Nevertheless for reversible engine transformation, the global unitary evolution strictly ensures total energy conservation of $B_1B_2S_{12}S_{W_{12}}$ and weighted-energy conservation of $B_1B_2$, and that are mathematically expressed by the commutation relations $\left[U, \ H_{B_1} + H_{B_2} + H_{S}  \right]=0$ and $\left[U, \ \beta_1  H_{B_1} + \beta_2  H_{B_2}  \right] =0$ respectively, where $H_{S}=H_{S_1}+H_{S_2}+H_{S_{W_{1}}}+H_{S_{W_{2}}}$ is the Hamiltonian of the system $S=S_{12}S_{W_{12}}$. 

\section{Conservation of weighted-energy implies conservation of entropy}
To understand the relationship between the conservation of entropy and the conservation of weighted-energy, we analyze the engine process in terms of the transformations happening in micro-canonical ensembles. The baths are assumed to be considerably large compared to the systems and the batteries. Since the global unitary implementation of the SLTOs strictly satisfy total energy conservation, we may concentrate on the transformation happening in each total energy block separately. For instance, consider the transformation \eqref{eq:E12Block}. This respects the total energy conservation, as given in Eq.~\eqref{eq:TotEngCons}. Again, the overall process occurs unitarily in isolation, so total entropy must be strictly conserved. The batteries only absorb or release work, and, by definition, they cannot exchange entropy with the rest of the system. Thus the entropy of system-bath composite ($B_1S_1S_2B_2$) must have to be conserved. Given the initial total energy $E_1 + E_2$, the strict entropy conservation implies the conservation of degeneracy, i.e.,
\begin{align}
g_B(E_1 + E_2)&=g_B(E_1^\prime + E_2^\prime) \nonumber \\
& = g_B(E_1 + E_2) e^{\beta_1 (E_1^\prime - E_1) + \beta_1 (E_2^\prime - E_2)}.
\end{align}
Thus, the following must have to be satisfied $\beta_1 (E_1^\prime - E_1) + \beta_1 (E_2^\prime - E_2) =0$. Here, $E_1=E_{B_1} + E_{S_1}$ and $E_2=E_{B_2} + E_{S_2}$, and similarly for $E_1^\prime$ and $E_2^\prime$. Thus, the above condition is reduced to 
\begin{align}
\beta_1 \left(\Delta E_{B_1} + \Delta E_{S_1}\right) + 	\beta_2 \left(\Delta E_{B_2} + \Delta E_{S_2}\right)=0,
\end{align}
where $\Delta E_x = E_x^\prime - E_x$ is the change in energy for the given total energy block. This is nothing but the condition for strict weighted-energy conservation, as ensured by the commutation relation~\eqref{eqmt:slto-commutationTemp}. 

\section{Engineering intensity-dependence in Hamiltonian~\eqref{eq:Hi}}
This section aims to show how the interaction Hamiltonian~\eqref{eq:Hi} can be realized with designed cavities or ion traps. We will focus here on the case of one cavity interacting with a two-level atom. Generalization to three-level systems and two different cavities coupled to the two different transitions is straightforward.

So, the starting point is a cavity (or trap) with a slight anharmonicity. That is described by the Hamiltonian of a harmonic oscillator, with frequency $\omega$ and a small controllable anharmonicity $V\left(x/x_0\right)$,
\begin{equation}
	H_{\rm cav} = \frac{p^2}{2m} +\frac{m\omega^2x^2}{2} + 
	V\left(x/x_0\right)= \hbar \omega \ a^\dag a + V\left(\frac{a^\dag+ a}{\sqrt{2}}\right).
\end{equation}
where $x_0=\sqrt{\frac{\hbar}{m\omega}}$.
Assuming that $\omega$ is much larger than any other relevant frequency, it makes sense to go to interaction picture with respect to the harmonic part of the Hamiltonian, and apply rotating wave approximation, i.e. neglect all rapidly oscillating terms and leave only diagonal terms in the Fock basis. The end result is
\begin{equation}
	H_{\rm cav,I} = f(N),
\end{equation}
where $N=a^\dag a$ and $f(n)= \langle n|V(x)|n\rangle$ with $N\ket{n}=n\ket{n}$.

Similarly, we assume that atom-cavity coupling originally has a general form
\begin{equation}
	H_{\rm c} = \hbar g b(x/x_{0})(\sigma^\dag + \sigma),
\end{equation}
where $b(x/x_{0})$ is the cavity mode function, which we take to be odd, i.e. $b(-y)=-b(y)$. Assuming that the atom is close to the bare cavity resonance and,  performing the same steps as before, we end up with the interaction Hamiltonian
\begin{equation}
	H_{\rm c,I} = \hbar g (\theta(N)\sigma^\dag a + h.c.),
\end{equation}
where $\langle n-1|b(x/x_{0})|n\rangle= \theta(n-1)\sqrt{n}$.

The intensity-dependent functions $f(N)$ and $\theta(N)$ are related to the original functions $V\left(x/x_0\right)$ and $b\left(x/x_0\right)$. For general  $f(N)$ and $\theta(N)$, one needs to design the original functions. This can be done using Monte Carlo (MC) optimization procedures. To this aim one defines a cost function
\begin{equation}
	C[V,b] = ||f_{\rm act}(\cdot)-f_{\rm tar}(\cdot)|| + ||\theta_{\rm act}(\cdot)-\theta_{\rm tar}(\cdot)||, 
\end{equation}
where $f,\theta_{\rm act,\rm tar}$ are the actual and target forms of the functions $f(\cdot)$ and $\theta(\cdot)$, and $||\cdot||$ denotes any norm in the space of the functions $f, \theta$. Judging from Eq.~\eqref{eq:ftheta} it can be $L^2$-norm for $f(\cdot)$ and $L^q$-norm with $q>2$ for $\theta(\cdot)$.
Now the MC procedure runs as follows: i) we choose actual form of $V_{\rm act}(\cdot)$ and $b_{\rm act}(\cdot)$; ii) we calculate $f_{\rm act}(\cdot)$, $\theta_{\rm act}(\cdot)$ and $C[V_{\rm act},b_{\rm act}]$; iii) we modify slightly $V_{\rm act}(\cdot)$ and $b_{\rm act}$ and calculate the new value of $C[V_{\rm act}, b_{\rm act}]$; iv) we accept the modification, if the new value of the error function is smaller than the previous one; v) we go to iii) and repeat this steps until convergence is achieved. MC optimization maybe modifies to allow small errors, if we treat the cost functions like energy and minimize the corresponding free energy at some arbitrary auxiliary temperature ${\cal T}$.

The question of the convergence of the MC procedure, as well as the sensitivity and the role of errors in the realization of our quantum engine is very interesting but clearly goes beyond the scope of present work. We will study it in a future publication. 

\section{Effective Hamiltonian of the quantum optics based quantum heat engine}
We start by considering that our system $S$ is described by the Hamiltonian of a $\Lambda$-system and that each of the two transitions is coupled to a different bosonic mode. The Hamiltonian is divided into two parts $H=H_0+H_1$, with
\begin{equation}
	\begin{split}
		H_0=\omega_1N_1&+\omega_2N_2 +E_1\ketbra{1}{1}_{S}  +E_2\ketbra{2}{2}_{S} +E_3\ketbra{3}{3}_{S}, \\
		H_1= f_1(N_1&)+f_2(N_2)+g_1\theta_1(N_1)(\aop\sigma_{31}+\text{h.c.}) \\
		&+g_2\theta_2(N_2)(\aaop\sigma_{32}+\text{h.c.}). \nonumber
	\end{split}
\end{equation}
Here $\sigma_{ij}=\ketbra{i}{j}_{S}$ is the transition operator, and  $N_k=a_k^\dagger a_k$ the number operator corresponding to the bath $B_k$. The system and bath energies are given by $E_i$ and $\omega_k$ ($\hbar=1$). The terms $f_k(N_k)$ represent intensity-dependent energy shifts of the baths, whereas $g_k(N_k)$ also takes into account the intensity-dependence of the dipole interaction between the system and the baths. We now set $E_1=0$ and move to the interaction picture with respect to $H_0'=\omega_1N_1+\omega_2N_2 +E_2\ketbra{2}{2}_{S} +\omega_1\ketbra{3}{3}_S$, imposing the resonant condition $\omega_1 =\omega_2 + E_2$. This gives the interaction picture Hamiltonian
\begin{align}
	H_I=  \Delta\ket{3}\bra{3}&+f_1(N_1)+f_2(N_2)+g_1\theta_1(N_1)(\aop\sigma_{31}+\text{h.c.}) \nonumber \\ &+g_2\theta_2(N_2)(\aaop\sigma_{32}+\text{h.c.}),
\end{align}
with $\Delta=E_3-\omega_1$.  We express our quantum state in the interaction picture as:
\begin{equation}
	\ket{\Psi} = \ket{\tilde{\alpha}}_B\ket{1}_S+\ket{\tilde{\beta}}_B\ket{2}_S+\ket{\tilde{\gamma}}_B\ket{3}_S,
\end{equation}
where $\ket{\tilde{\alpha}}_B,\ \ket{\tilde{\beta}}_B$, and $\ket{\tilde{\gamma}}_B$ are unnormalized states of the $B_1B_2$ composite. This leads to the following form of the Schr\"odinger equation in components
\begin{equation}
	i\frac{\text{d}}{\text{dt}}\ket{\tilde{\alpha}}_B=\left[ f_1(N_1)+f_2(N_2)\right]\ket{\tilde{\alpha}}_B+g_1\theta_1(N_1)\adop\ket{\tilde{\gamma}}_B,
\end{equation}
\begin{equation}
	i\frac{\text{d}}{\text{dt}}\ket{\tilde{\beta}}_B=\left[f_1(N_1)+f_2(N_2)\right]\ket{\tilde{\beta}}_B+g_2\theta_2(N_2)\aadop\ket{\tilde{\gamma}}_B,
\end{equation}
\begin{align}
	i\frac{\text{d}}{\text{dt}}\ket{\tilde{\gamma}}_B=&\left[f_1(N_1)+f_2(N_2)+\Delta\right]\ket{\tilde{\gamma}}_B \nonumber \\ &+g_2\theta_2(N_2)\aaop\ket{\tilde{\beta}}_B+g_1\theta_1(N_1)\aop\ket{\tilde{\alpha}}_B.
\end{align}
Now we consider a large detuning , i.e., $\norm{\frac{\text{d}}{\text{dt}}\ket{\tilde{\gamma}}_B}\simeq 0$, and $\Delta \gg \mean{f_1(N_1)+f_2(N_2)}$, which allows us to express

\begin{equation}
	\ket{\tilde{\gamma}}\simeq -\frac{1}{\Delta}\left(g_1\theta_1(N_1)\aop\ket{\tilde{\alpha}}_B+g_2\theta_2(N_2)\aaop\ket{\tilde{\beta}}_B\right).
\end{equation}
Introducing this result in the previous equations leads to 
\begin{align}
	i\frac{\text{d}}{\text{dt}}\ket{\tilde{\alpha}}_B=&\left[ f_1(N_1)+f_2(N_2)-\frac{g_1^2}{\Delta}\theta_1^2(N_1)\right]\ket{\tilde{\alpha}}_B \nonumber \\ &-\frac{g_1g_2}{\Delta}\theta_1(N_1)\adop\aaop\theta_2(N_2)\ket{\tilde{\beta}}_B,
\end{align}
\begin{align}
	i\frac{\text{d}}{\text{dt}}\ket{\tilde{\beta}}_B=&\left[f_1(N_1)+f_2(N_2)-\frac{g_2^2}{\Delta}\theta_2^2(N_2)\right]\ket{\tilde{\beta}}_B \nonumber \\ &-\frac{g_1g_2}{\Delta}\theta_2(N_2)\aadop\aop\theta_1(N_1)\ket{\tilde{\alpha}}_B.
\end{align}
The latter are the same equations of motion generated by an effective interacting Hamiltonian given by
\begin{align}
	H'_\text{eff}=&f_1(N_1)+f_2(N_2)-\frac{g_1^2}{\Delta}\theta_1^2(N_1)-\frac{g_2^2}{\Delta}\theta_2^2(N_2) \nonumber \\ &-\frac{g_1g_2}{\Delta}(\theta_1(N_1)\adop\aaop\theta_2(N_2)\sigma_{21}+\text{h.c.}).
\end{align}
For suitable functions that satisfy
\begin{equation}
	f_k(N_k)=\frac{g_k^2}{\Delta}\theta_k^2(N_k),
\end{equation}
we obtain the final effective Hamiltonian
\begin{equation}
	H_\text{eff}=-\frac{g_1g_2}{\Delta}\theta_1(N_1)\adop\aaop\theta_2(N_2)\sigma_{21}+\text{h.c.}.
\end{equation}


\begin{thebibliography}{58}%
	\makeatletter
	\providecommand \@ifxundefined [1]{%
		\@ifx{#1\undefined}
	}%
	\providecommand \@ifnum [1]{%
		\ifnum #1\expandafter \@firstoftwo
		\else \expandafter \@secondoftwo
		\fi
	}%
	\providecommand \@ifx [1]{%
		\ifx #1\expandafter \@firstoftwo
		\else \expandafter \@secondoftwo
		\fi
	}%
	\providecommand \natexlab [1]{#1}%
	\providecommand \enquote  [1]{``#1''}%
	\providecommand \bibnamefont  [1]{#1}%
	\providecommand \bibfnamefont [1]{#1}%
	\providecommand \citenamefont [1]{#1}%
	\providecommand \href@noop [0]{\@secondoftwo}%
	\providecommand \href [0]{\begingroup \@sanitize@url \@href}%
	\providecommand \@href[1]{\@@startlink{#1}\@@href}%
	\providecommand \@@href[1]{\endgroup#1\@@endlink}%
	\providecommand \@sanitize@url [0]{\catcode `\\12\catcode `\$12\catcode
		`\&12\catcode `\#12\catcode `\^12\catcode `\_12\catcode `\%12\relax}%
	\providecommand \@@startlink[1]{}%
	\providecommand \@@endlink[0]{}%
	\providecommand \url  [0]{\begingroup\@sanitize@url \@url }%
	\providecommand \@url [1]{\endgroup\@href {#1}{\urlprefix }}%
	\providecommand \urlprefix  [0]{URL }%
	\providecommand \Eprint [0]{\href }%
	\providecommand \doibase [0]{http://dx.doi.org/}%
	\providecommand \selectlanguage [0]{\@gobble}%
	\providecommand \bibinfo  [0]{\@secondoftwo}%
	\providecommand \bibfield  [0]{\@secondoftwo}%
	\providecommand \translation [1]{[#1]}%
	\providecommand \BibitemOpen [0]{}%
	\providecommand \bibitemStop [0]{}%
	\providecommand \bibitemNoStop [0]{.\EOS\space}%
	\providecommand \EOS [0]{\spacefactor3000\relax}%
	\providecommand \BibitemShut  [1]{\csname bibitem#1\endcsname}%
	\let\auto@bib@innerbib\@empty
	\bibitem [{\citenamefont {Curzon}\ and\ \citenamefont
		{Ahlborn}(1975)}]{Curzon75}%
	\BibitemOpen
	\bibfield  {author} {\bibinfo {author} {\bibfnamefont {F.~L.}\ \bibnamefont
			{Curzon}}\ and\ \bibinfo {author} {\bibfnamefont {B.}~\bibnamefont
			{Ahlborn}},\ }\bibfield  {title} {\enquote {\bibinfo {title} {Efficiency of a
				carnot engine at maximum power output},}\ }\href {\doibase 10.1119/1.10023}
	{\bibfield  {journal} {\bibinfo  {journal} {American Journal of Physics}\
		}\textbf {\bibinfo {volume} {43}},\ \bibinfo {pages} {22--24} (\bibinfo
		{year} {1975})}\BibitemShut {NoStop}%
	\bibitem [{\citenamefont {Berry}\ \emph {et~al.}(2000)\citenamefont {Berry},
		\citenamefont {Kazakov}, \citenamefont {Sieniutycz}, \citenamefont {Szwast},\
		and\ \citenamefont {Tsirlin}}]{Berry00}%
	\BibitemOpen
	\bibfield  {author} {\bibinfo {author} {\bibfnamefont {R.~S.}\ \bibnamefont
			{Berry}}, \bibinfo {author} {\bibfnamefont {V.}~\bibnamefont {Kazakov}},
		\bibinfo {author} {\bibfnamefont {S.}~\bibnamefont {Sieniutycz}}, \bibinfo
		{author} {\bibfnamefont {Z.}~\bibnamefont {Szwast}}, \ and\ \bibinfo {author}
		{\bibfnamefont {A.~M.}\ \bibnamefont {Tsirlin}},\ }\href
	{https://www.wiley.com/en-us/Thermodynamic+Optimization+of+Finite+Time+Processes-p-9780471967521}
	{\emph {\bibinfo {title} {Thermodynamic Optimization of Finite-Time
				Processes}}}\ (\bibinfo  {publisher} {Wiley},\ \bibinfo {year}
	{2000})\BibitemShut {NoStop}%
	\bibitem [{\citenamefont {Salamon}\ \emph {et~al.}(2001)\citenamefont
		{Salamon}, \citenamefont {Nulton}, \citenamefont {Siragusa}, \citenamefont
		{Andersen},\ and\ \citenamefont {Limon}}]{Salamon01}%
	\BibitemOpen
	\bibfield  {author} {\bibinfo {author} {\bibfnamefont {P.}~\bibnamefont
			{Salamon}}, \bibinfo {author} {\bibfnamefont {J.~D.}\ \bibnamefont {Nulton}},
		\bibinfo {author} {\bibfnamefont {G.}~\bibnamefont {Siragusa}}, \bibinfo
		{author} {\bibfnamefont {T.~R.}\ \bibnamefont {Andersen}}, \ and\ \bibinfo
		{author} {\bibfnamefont {A.}~\bibnamefont {Limon}},\ }\bibfield  {title}
	{\enquote {\bibinfo {title} {Principles of control thermodynamics},}\ }\href
	{\doibase https://doi.org/10.1016/S0360-5442(00)00059-1} {\bibfield
		{journal} {\bibinfo  {journal} {Energy}\ }\textbf {\bibinfo {volume} {26}},\
		\bibinfo {pages} {307 -- 319} (\bibinfo {year} {2001})}\BibitemShut {NoStop}%
	\bibitem [{\citenamefont {Binder}\ \emph {et~al.}(2018)\citenamefont {Binder},
		\citenamefont {Correa}, \citenamefont {Gogolin}, \citenamefont {Anders},\
		and\ \citenamefont {Adesso}}]{Binder18}%
	\BibitemOpen
	\bibfield  {author} {\bibinfo {author} {\bibfnamefont {Felix}\ \bibnamefont
			{Binder}}, \bibinfo {author} {\bibfnamefont {Luis~A.}\ \bibnamefont
			{Correa}}, \bibinfo {author} {\bibfnamefont {Christian}\ \bibnamefont
			{Gogolin}}, \bibinfo {author} {\bibfnamefont {Janet}\ \bibnamefont {Anders}},
		\ and\ \bibinfo {author} {\bibfnamefont {Gerardo}\ \bibnamefont {Adesso}},\
	}\href {\doibase 10.1007/978-3-319-99046-0} {\emph {\bibinfo {title}
			{Thermodynamics in the Quantum Regime}}},\ Vol.\ \bibinfo {volume} {195}\
	(\bibinfo  {publisher} {Springer International Publishing},\ \bibinfo {year}
	{2018})\BibitemShut {NoStop}%
	\bibitem [{\citenamefont {Jarzynski}(1997)}]{Jarzynski97}%
	\BibitemOpen
	\bibfield  {author} {\bibinfo {author} {\bibfnamefont {C.}~\bibnamefont
			{Jarzynski}},\ }\bibfield  {title} {\enquote {\bibinfo {title}
			{Nonequilibrium equality for free energy differences},}\ }\href {\doibase
		10.1103/PhysRevLett.78.2690} {\bibfield  {journal} {\bibinfo  {journal}
			{Phys. Rev. Lett.}\ }\textbf {\bibinfo {volume} {78}},\ \bibinfo {pages}
		{2690--2693} (\bibinfo {year} {1997})}\BibitemShut {NoStop}%
	\bibitem [{\citenamefont {Crooks}(1999)}]{Crooks99}%
	\BibitemOpen
	\bibfield  {author} {\bibinfo {author} {\bibfnamefont {Gavin~E.}\
			\bibnamefont {Crooks}},\ }\bibfield  {title} {\enquote {\bibinfo {title}
			{Entropy production fluctuation theorem and the nonequilibrium work relation
				for free energy differences},}\ }\href {\doibase 10.1103/PhysRevE.60.2721}
	{\bibfield  {journal} {\bibinfo  {journal} {Phys. Rev. E}\ }\textbf {\bibinfo
			{volume} {60}},\ \bibinfo {pages} {2721--2726} (\bibinfo {year}
		{1999})}\BibitemShut {NoStop}%
	\bibitem [{\citenamefont {Campisi}\ \emph {et~al.}(2011)\citenamefont
		{Campisi}, \citenamefont {H\"anggi},\ and\ \citenamefont
		{Talkner}}]{Campisi11}%
	\BibitemOpen
	\bibfield  {author} {\bibinfo {author} {\bibfnamefont {Michele}\ \bibnamefont
			{Campisi}}, \bibinfo {author} {\bibfnamefont {Peter}\ \bibnamefont
			{H\"anggi}}, \ and\ \bibinfo {author} {\bibfnamefont {Peter}\ \bibnamefont
			{Talkner}},\ }\bibfield  {title} {\enquote {\bibinfo {title} {Colloquium:
				Quantum fluctuation relations: Foundations and applications},}\ }\href
	{\doibase 10.1103/RevModPhys.83.771} {\bibfield  {journal} {\bibinfo
			{journal} {Rev. Mod. Phys.}\ }\textbf {\bibinfo {volume} {83}},\ \bibinfo
		{pages} {771--791} (\bibinfo {year} {2011})}\BibitemShut {NoStop}%
	\bibitem [{\citenamefont {Alhambra}\ \emph {et~al.}(2016)\citenamefont
		{Alhambra}, \citenamefont {Masanes}, \citenamefont {Oppenheim},\ and\
		\citenamefont {Perry}}]{Alhambra16}%
	\BibitemOpen
	\bibfield  {author} {\bibinfo {author} {\bibfnamefont {\'Alvaro~M.}\
			\bibnamefont {Alhambra}}, \bibinfo {author} {\bibfnamefont {Lluis}\
			\bibnamefont {Masanes}}, \bibinfo {author} {\bibfnamefont {Jonathan}\
			\bibnamefont {Oppenheim}}, \ and\ \bibinfo {author} {\bibfnamefont
			{Christopher}\ \bibnamefont {Perry}},\ }\bibfield  {title} {\enquote
		{\bibinfo {title} {Fluctuating work: From quantum thermodynamical identities
				to a second law equality},}\ }\href {\doibase 10.1103/PhysRevX.6.041017}
	{\bibfield  {journal} {\bibinfo  {journal} {Phys. Rev. X}\ }\textbf {\bibinfo
			{volume} {6}},\ \bibinfo {pages} {041017} (\bibinfo {year}
		{2016})}\BibitemShut {NoStop}%
	\bibitem [{\citenamefont {\AA{}berg}(2018)}]{Aberg18}%
	\BibitemOpen
	\bibfield  {author} {\bibinfo {author} {\bibfnamefont {Johan}\ \bibnamefont
			{\AA{}berg}},\ }\bibfield  {title} {\enquote {\bibinfo {title} {Fully quantum
				fluctuation theorems},}\ }\href {\doibase 10.1103/PhysRevX.8.011019}
	{\bibfield  {journal} {\bibinfo  {journal} {Phys. Rev. X}\ }\textbf {\bibinfo
			{volume} {8}},\ \bibinfo {pages} {011019} (\bibinfo {year}
		{2018})}\BibitemShut {NoStop}%
	\bibitem [{\citenamefont {Brandao}\ \emph {et~al.}(2013)\citenamefont
		{Brandao}, \citenamefont {Horodecki}, \citenamefont {Oppenheim},
		\citenamefont {Renes},\ and\ \citenamefont {Spekkens}}]{Brandao13}%
	\BibitemOpen
	\bibfield  {author} {\bibinfo {author} {\bibfnamefont {Fernando G. S.~L.}\
			\bibnamefont {Brandao}}, \bibinfo {author} {\bibfnamefont {Michal}\
			\bibnamefont {Horodecki}}, \bibinfo {author} {\bibfnamefont {Jonathan}\
			\bibnamefont {Oppenheim}}, \bibinfo {author} {\bibfnamefont {Joseph~M.}\
			\bibnamefont {Renes}}, \ and\ \bibinfo {author} {\bibfnamefont {Robert~W.}\
			\bibnamefont {Spekkens}},\ }\bibfield  {title} {\enquote {\bibinfo {title}
			{Resource theory of quantum states out of thermal equilibrium},}\ }\href
	{\doibase 10.1103/PhysRevLett.111.250404} {\bibfield  {journal} {\bibinfo
			{journal} {Phys. Rev. Lett.}\ }\textbf {\bibinfo {volume} {111}},\ \bibinfo
		{pages} {250404} (\bibinfo {year} {2013})}\BibitemShut {NoStop}%
	\bibitem [{\citenamefont {Horodecki}\ and\ \citenamefont
		{Oppenheim}(2013)}]{Horodecki13}%
	\BibitemOpen
	\bibfield  {author} {\bibinfo {author} {\bibfnamefont {Michal}\ \bibnamefont
			{Horodecki}}\ and\ \bibinfo {author} {\bibfnamefont {Jonathan}\ \bibnamefont
			{Oppenheim}},\ }\bibfield  {title} {\enquote {\bibinfo {title} {Fundamental
				limitations for quantum and nanoscale thermodynamics},}\ }\href {\doibase
		10.1038/ncomms3059} {\bibfield  {journal} {\bibinfo  {journal} {Nat.
				Commun.}\ }\textbf {\bibinfo {volume} {4}},\ \bibinfo {pages} {2059}
		(\bibinfo {year} {2013})}\BibitemShut {NoStop}%
	\bibitem [{\citenamefont {Skrzypczyk}\ \emph {et~al.}(2014)\citenamefont
		{Skrzypczyk}, \citenamefont {Short},\ and\ \citenamefont
		{Popescu}}]{Skrzypczyk14}%
	\BibitemOpen
	\bibfield  {author} {\bibinfo {author} {\bibfnamefont {Paul}\ \bibnamefont
			{Skrzypczyk}}, \bibinfo {author} {\bibfnamefont {Anthony~J.}\ \bibnamefont
			{Short}}, \ and\ \bibinfo {author} {\bibfnamefont {Sandu}\ \bibnamefont
			{Popescu}},\ }\bibfield  {title} {\enquote {\bibinfo {title} {Work extraction
				and thermodynamics for individual quantum systems},}\ }\href {\doibase
		10.1038/ncomms5185} {\bibfield  {journal} {\bibinfo  {journal} {Nat.
				Commun.}\ }\textbf {\bibinfo {volume} {5}},\ \bibinfo {pages} {4185}
		(\bibinfo {year} {2014})}\BibitemShut {NoStop}%
	\bibitem [{\citenamefont {Brandao}\ \emph {et~al.}(2015)\citenamefont
		{Brandao}, \citenamefont {Horodecki}, \citenamefont {Ng}, \citenamefont
		{Oppenheim},\ and\ \citenamefont {Wehner}}]{Brandao15}%
	\BibitemOpen
	\bibfield  {author} {\bibinfo {author} {\bibfnamefont {Fernando G. S.~L.}\
			\bibnamefont {Brandao}}, \bibinfo {author} {\bibfnamefont {Michal}\
			\bibnamefont {Horodecki}}, \bibinfo {author} {\bibfnamefont {Nelly}\
			\bibnamefont {Ng}}, \bibinfo {author} {\bibfnamefont {Jonathan}\ \bibnamefont
			{Oppenheim}}, \ and\ \bibinfo {author} {\bibfnamefont {Stephanie}\
			\bibnamefont {Wehner}},\ }\bibfield  {title} {\enquote {\bibinfo {title} {The
				second laws of quantum thermodynamics},}\ }\href {\doibase
		doi:10.1073/pnas.1411728112} {\bibfield  {journal} {\bibinfo  {journal}
			{Proc. Natl. Acad. Sci.}\ }\textbf {\bibinfo {volume} {112}},\ \bibinfo
		{pages} {3275--3279} (\bibinfo {year} {2015})}\BibitemShut {NoStop}%
	\bibitem [{\citenamefont {Lostaglio}\ \emph {et~al.}(2015)\citenamefont
		{Lostaglio}, \citenamefont {Jennings},\ and\ \citenamefont
		{Rudolph}}]{Lostaglio15a}%
	\BibitemOpen
	\bibfield  {author} {\bibinfo {author} {\bibfnamefont {Matteo}\ \bibnamefont
			{Lostaglio}}, \bibinfo {author} {\bibfnamefont {David}\ \bibnamefont
			{Jennings}}, \ and\ \bibinfo {author} {\bibfnamefont {Terry}\ \bibnamefont
			{Rudolph}},\ }\bibfield  {title} {\enquote {\bibinfo {title} {Description of
				quantum coherence in thermodynamic processes requires constraints beyond free
				energy},}\ }\href {\doibase 10.1038/ncomms7383} {\bibfield  {journal}
		{\bibinfo  {journal} {Nat. Commun.}\ }\textbf {\bibinfo {volume} {6}},\
		\bibinfo {pages} {6383} (\bibinfo {year} {2015})}\BibitemShut {NoStop}%
	\bibitem [{\citenamefont {Bera}\ \emph {et~al.}(2017)\citenamefont {Bera},
		\citenamefont {Riera}, \citenamefont {Lewenstein},\ and\ \citenamefont
		{Winter}}]{Bera16}%
	\BibitemOpen
	\bibfield  {author} {\bibinfo {author} {\bibfnamefont {Manabendra~Nath}\
			\bibnamefont {Bera}}, \bibinfo {author} {\bibfnamefont {Arnau}\ \bibnamefont
			{Riera}}, \bibinfo {author} {\bibfnamefont {Maciej}\ \bibnamefont
			{Lewenstein}}, \ and\ \bibinfo {author} {\bibfnamefont {Andreas}\
			\bibnamefont {Winter}},\ }\bibfield  {title} {\enquote {\bibinfo {title}
			{Generalized laws of thermodynamics in the presence of correlations},}\
	}\href {\doibase doi:10.1038/s41467-017-02370-x} {\bibfield  {journal}
		{\bibinfo  {journal} {Nat. Commun.}\ }\textbf {\bibinfo {volume} {8}},\
		\bibinfo {pages} {2180} (\bibinfo {year} {2017})}\BibitemShut {NoStop}%
	\bibitem [{\citenamefont {Gour}\ \emph {et~al.}(2018)\citenamefont {Gour},
		\citenamefont {Jennings}, \citenamefont {Buscemi},\ and\ \citenamefont
		{Duan}}]{Gour2018}%
	\BibitemOpen
	\bibfield  {author} {\bibinfo {author} {\bibfnamefont {Gilad}\ \bibnamefont
			{Gour}}, \bibinfo {author} {\bibfnamefont {David}\ \bibnamefont {Jennings}},
		\bibinfo {author} {\bibfnamefont {Francesco}\ \bibnamefont {Buscemi}}, \ and\
		\bibinfo {author} {\bibfnamefont {Iman}\ \bibnamefont {Duan}, \bibfnamefont
			{Runyao~Marvian}},\ }\bibfield  {title} {\enquote {\bibinfo {title} {Quantum
				majorization and a complete set of entropic conditions for quantum
				thermodynamics},}\ }\href {\doibase 10.1038/s41467-018-06261-7} {\bibfield
		{journal} {\bibinfo  {journal} {Nat. Commun.}\ }\textbf {\bibinfo {volume}
			{9}},\ \bibinfo {pages} {5352} (\bibinfo {year} {2018})}\BibitemShut
	{NoStop}%
	\bibitem [{\citenamefont {M\"uller}(2018)}]{Muller18}%
	\BibitemOpen
	\bibfield  {author} {\bibinfo {author} {\bibfnamefont {Markus~P.}\
			\bibnamefont {M\"uller}},\ }\bibfield  {title} {\enquote {\bibinfo {title}
			{Correlating thermal machines and the second law at the nanoscale},}\ }\href
	{\doibase 10.1103/PhysRevX.8.041051} {\bibfield  {journal} {\bibinfo
			{journal} {Phys. Rev. X}\ }\textbf {\bibinfo {volume} {8}},\ \bibinfo {pages}
		{041051} (\bibinfo {year} {2018})}\BibitemShut {NoStop}%
	\bibitem [{\citenamefont {Sparaciari}\ \emph {et~al.}(2017)\citenamefont
		{Sparaciari}, \citenamefont {Oppenheim},\ and\ \citenamefont
		{Fritz}}]{Sparaciari17}%
	\BibitemOpen
	\bibfield  {author} {\bibinfo {author} {\bibfnamefont {Carlo}\ \bibnamefont
			{Sparaciari}}, \bibinfo {author} {\bibfnamefont {Jonathan}\ \bibnamefont
			{Oppenheim}}, \ and\ \bibinfo {author} {\bibfnamefont {Tobias}\ \bibnamefont
			{Fritz}},\ }\bibfield  {title} {\enquote {\bibinfo {title} {Resource theory
				for work and heat},}\ }\href {\doibase 10.1103/PhysRevA.96.052112} {\bibfield
		{journal} {\bibinfo  {journal} {Phys. Rev. A}\ }\textbf {\bibinfo {volume}
			{96}},\ \bibinfo {pages} {052112} (\bibinfo {year} {2017})}\BibitemShut
	{NoStop}%
	\bibitem [{\citenamefont {Bera}\ \emph {et~al.}(2019)\citenamefont {Bera},
		\citenamefont {Riera}, \citenamefont {Lewenstein}, \citenamefont {Khanian},\
		and\ \citenamefont {Winter}}]{Bera17}%
	\BibitemOpen
	\bibfield  {author} {\bibinfo {author} {\bibfnamefont {Manabendra~Nath}\
			\bibnamefont {Bera}}, \bibinfo {author} {\bibfnamefont {Arnau}\ \bibnamefont
			{Riera}}, \bibinfo {author} {\bibfnamefont {Maciej}\ \bibnamefont
			{Lewenstein}}, \bibinfo {author} {\bibfnamefont {Zahra~Baghali}\ \bibnamefont
			{Khanian}}, \ and\ \bibinfo {author} {\bibfnamefont {Andreas}\ \bibnamefont
			{Winter}},\ }\bibfield  {title} {\enquote {\bibinfo {title} {Thermodynamics
				as a {C}onsequence of {I}nformation {C}onservation},}\ }\href {\doibase
		10.22331/q-2019-02-14-121} {\bibfield  {journal} {\bibinfo  {journal}
			{{Quantum}}\ }\textbf {\bibinfo {volume} {3}},\ \bibinfo {pages} {121}
		(\bibinfo {year} {2019})}\BibitemShut {NoStop}%
	\bibitem [{\citenamefont {Uzdin}\ and\ \citenamefont {Rahav}(2018)}]{Uzdin18}%
	\BibitemOpen
	\bibfield  {author} {\bibinfo {author} {\bibfnamefont {Raam}\ \bibnamefont
			{Uzdin}}\ and\ \bibinfo {author} {\bibfnamefont {Saar}\ \bibnamefont
			{Rahav}},\ }\bibfield  {title} {\enquote {\bibinfo {title} {Global passivity
				in microscopic thermodynamics},}\ }\href {\doibase 10.1103/PhysRevX.8.021064}
	{\bibfield  {journal} {\bibinfo  {journal} {Phys. Rev. X}\ }\textbf {\bibinfo
			{volume} {8}},\ \bibinfo {pages} {021064} (\bibinfo {year}
		{2018})}\BibitemShut {NoStop}%
	\bibitem [{\citenamefont {Baghali~Khanian}\ \emph {et~al.}(2020)\citenamefont
		{Baghali~Khanian}, \citenamefont {Bera}, \citenamefont {Riera}, \citenamefont
		{Lewenstein},\ and\ \citenamefont {Winter}}]{Khanian20}%
	\BibitemOpen
	\bibfield  {author} {\bibinfo {author} {\bibfnamefont {Zahra}\ \bibnamefont
			{Baghali~Khanian}}, \bibinfo {author} {\bibfnamefont {Manabendra~Nath}\
			\bibnamefont {Bera}}, \bibinfo {author} {\bibfnamefont {Arnau}\ \bibnamefont
			{Riera}}, \bibinfo {author} {\bibfnamefont {Maciej}\ \bibnamefont
			{Lewenstein}}, \ and\ \bibinfo {author} {\bibfnamefont {Andreas}\
			\bibnamefont {Winter}},\ }\bibfield  {title} {\enquote {\bibinfo {title}
			{Resource theory of heat and work with non-commuting charges: yet another new
				foundation of thermodynamics},}\ }\href {https://arxiv.org/abs/2011.08020}
	{\bibfield  {journal} {\bibinfo  {journal} {arXiv:2011.08020}\ } (\bibinfo
		{year} {2020})}\BibitemShut {NoStop}%
	\bibitem [{\citenamefont {Kosloff}\ and\ \citenamefont
		{Levy}(2014)}]{Kosloff14}%
	\BibitemOpen
	\bibfield  {author} {\bibinfo {author} {\bibfnamefont {Ronnie}\ \bibnamefont
			{Kosloff}}\ and\ \bibinfo {author} {\bibfnamefont {Amikam}\ \bibnamefont
			{Levy}},\ }\bibfield  {title} {\enquote {\bibinfo {title} {Quantum heat
				engines and refrigerators: Continuous devices},}\ }\href {\doibase
		10.1146/annurev-physchem-040513-103724} {\bibfield  {journal} {\bibinfo
			{journal} {Annu. Rev. Phys. Chem.}\ }\textbf {\bibinfo {volume} {65}},\
		\bibinfo {pages} {365--393} (\bibinfo {year} {2014})}\BibitemShut {NoStop}%
	\bibitem [{\citenamefont {Uzdin}\ \emph {et~al.}(2015)\citenamefont {Uzdin},
		\citenamefont {Levy},\ and\ \citenamefont {Kosloff}}]{Uzdin15}%
	\BibitemOpen
	\bibfield  {author} {\bibinfo {author} {\bibfnamefont {Raam}\ \bibnamefont
			{Uzdin}}, \bibinfo {author} {\bibfnamefont {Amikam}\ \bibnamefont {Levy}}, \
		and\ \bibinfo {author} {\bibfnamefont {Ronnie}\ \bibnamefont {Kosloff}},\
	}\bibfield  {title} {\enquote {\bibinfo {title} {Equivalence of quantum heat
				machines, and quantum-thermodynamic signatures},}\ }\href {\doibase
		10.1103/PhysRevX.5.031044} {\bibfield  {journal} {\bibinfo  {journal} {Phys.
				Rev. X}\ }\textbf {\bibinfo {volume} {5}},\ \bibinfo {pages} {031044}
		(\bibinfo {year} {2015})}\BibitemShut {NoStop}%
	\bibitem [{\citenamefont {Klaers}\ \emph {et~al.}(2017)\citenamefont {Klaers},
		\citenamefont {Faelt}, \citenamefont {Imamoglu},\ and\ \citenamefont
		{Togan}}]{Klaers17}%
	\BibitemOpen
	\bibfield  {author} {\bibinfo {author} {\bibfnamefont {Jan}\ \bibnamefont
			{Klaers}}, \bibinfo {author} {\bibfnamefont {Stefan}\ \bibnamefont {Faelt}},
		\bibinfo {author} {\bibfnamefont {Atac}\ \bibnamefont {Imamoglu}}, \ and\
		\bibinfo {author} {\bibfnamefont {Emre}\ \bibnamefont {Togan}},\ }\bibfield
	{title} {\enquote {\bibinfo {title} {Squeezed thermal reservoirs as a
				resource for a nanomechanical engine beyond the carnot limit},}\ }\href
	{\doibase 10.1103/PhysRevX.7.031044} {\bibfield  {journal} {\bibinfo
			{journal} {Phys. Rev. X}\ }\textbf {\bibinfo {volume} {7}},\ \bibinfo {pages}
		{031044} (\bibinfo {year} {2017})}\BibitemShut {NoStop}%
	\bibitem [{\citenamefont {Ro\ss{}nagel}\ \emph {et~al.}(2014)\citenamefont
		{Ro\ss{}nagel}, \citenamefont {Abah}, \citenamefont {Schmidt-Kaler},
		\citenamefont {Singer},\ and\ \citenamefont {Lutz}}]{Rossnagel14}%
	\BibitemOpen
	\bibfield  {author} {\bibinfo {author} {\bibfnamefont {J.}~\bibnamefont
			{Ro\ss{}nagel}}, \bibinfo {author} {\bibfnamefont {O.}~\bibnamefont {Abah}},
		\bibinfo {author} {\bibfnamefont {F.}~\bibnamefont {Schmidt-Kaler}}, \bibinfo
		{author} {\bibfnamefont {K.}~\bibnamefont {Singer}}, \ and\ \bibinfo {author}
		{\bibfnamefont {E.}~\bibnamefont {Lutz}},\ }\bibfield  {title} {\enquote
		{\bibinfo {title} {Nanoscale heat engine beyond the carnot limit},}\ }\href
	{\doibase 10.1103/PhysRevLett.112.030602} {\bibfield  {journal} {\bibinfo
			{journal} {Phys. Rev. Lett.}\ }\textbf {\bibinfo {volume} {112}},\ \bibinfo
		{pages} {030602} (\bibinfo {year} {2014})}\BibitemShut {NoStop}%
	\bibitem [{\citenamefont {Verley}\ \emph {et~al.}(2014)\citenamefont {Verley},
		\citenamefont {Esposito}, \citenamefont {Willaert},\ and\ \citenamefont
		{Van~den Broeck}}]{Verley14}%
	\BibitemOpen
	\bibfield  {author} {\bibinfo {author} {\bibfnamefont {Gatien}\ \bibnamefont
			{Verley}}, \bibinfo {author} {\bibfnamefont {Massimiliano}\ \bibnamefont
			{Esposito}}, \bibinfo {author} {\bibfnamefont {Tim}\ \bibnamefont
			{Willaert}}, \ and\ \bibinfo {author} {\bibfnamefont {Christian}\
			\bibnamefont {Van~den Broeck}},\ }\bibfield  {title} {\enquote {\bibinfo
			{title} {The unlikely carnot efficiency},}\ }\href {\doibase
		10.1038/ncomms5721} {\bibfield  {journal} {\bibinfo  {journal} {Nat.
				Commun.}\ }\textbf {\bibinfo {volume} {5}},\ \bibinfo {pages} {4721}
		(\bibinfo {year} {2014})}\BibitemShut {NoStop}%
	\bibitem [{\citenamefont {Funo}\ and\ \citenamefont {Ueda}(2015)}]{Funo15}%
	\BibitemOpen
	\bibfield  {author} {\bibinfo {author} {\bibfnamefont {Ken}\ \bibnamefont
			{Funo}}\ and\ \bibinfo {author} {\bibfnamefont {Masahito}\ \bibnamefont
			{Ueda}},\ }\bibfield  {title} {\enquote {\bibinfo {title} {Work
				fluctuation-dissipation trade-off in heat engines},}\ }\href {\doibase
		10.1103/PhysRevLett.115.260601} {\bibfield  {journal} {\bibinfo  {journal}
			{Phys. Rev. Lett.}\ }\textbf {\bibinfo {volume} {115}},\ \bibinfo {pages}
		{260601} (\bibinfo {year} {2015})}\BibitemShut {NoStop}%
	\bibitem [{\citenamefont {Ng}\ \emph {et~al.}(2017)\citenamefont {Ng},
		\citenamefont {Woods},\ and\ \citenamefont {Wehner}}]{Ng17}%
	\BibitemOpen
	\bibfield  {author} {\bibinfo {author} {\bibfnamefont {Nelly Huei~Ying}\
			\bibnamefont {Ng}}, \bibinfo {author} {\bibfnamefont {Mischa~Prebin}\
			\bibnamefont {Woods}}, \ and\ \bibinfo {author} {\bibfnamefont {Stephanie}\
			\bibnamefont {Wehner}},\ }\bibfield  {title} {\enquote {\bibinfo {title}
			{Surpassing the carnot efficiency by extracting imperfect work},}\ }\href
	{\doibase 10.1088/1367-2630/aa8ced} {\bibfield  {journal} {\bibinfo
			{journal} {New J. Phys.}\ }\textbf {\bibinfo {volume} {19}},\ \bibinfo
		{pages} {113005} (\bibinfo {year} {2017})}\BibitemShut {NoStop}%
	\bibitem [{\citenamefont {Woods}\ \emph {et~al.}(2019)\citenamefont {Woods},
		\citenamefont {Ng},\ and\ \citenamefont {Wehner}}]{Woods19}%
	\BibitemOpen
	\bibfield  {author} {\bibinfo {author} {\bibfnamefont {Mischa~P.}\
			\bibnamefont {Woods}}, \bibinfo {author} {\bibfnamefont {Nelly Huei~Ying}\
			\bibnamefont {Ng}}, \ and\ \bibinfo {author} {\bibfnamefont {Stephanie}\
			\bibnamefont {Wehner}},\ }\bibfield  {title} {\enquote {\bibinfo {title} {The
				maximum efficiency of nano heat engines depends on more than temperature},}\
	}\href {\doibase 10.22331/q-2019-08-19-177} {\bibfield  {journal} {\bibinfo
			{journal} {{Quantum}}\ }\textbf {\bibinfo {volume} {3}},\ \bibinfo {pages}
		{177} (\bibinfo {year} {2019})}\BibitemShut {NoStop}%
	\bibitem [{\citenamefont {Manikandan}\ \emph {et~al.}(2019)\citenamefont
		{Manikandan}, \citenamefont {Dabelow}, \citenamefont {Eichhorn},\ and\
		\citenamefont {Krishnamurthy}}]{Manikandan19}%
	\BibitemOpen
	\bibfield  {author} {\bibinfo {author} {\bibfnamefont {Sreekanth~K.}\
			\bibnamefont {Manikandan}}, \bibinfo {author} {\bibfnamefont {Lennart}\
			\bibnamefont {Dabelow}}, \bibinfo {author} {\bibfnamefont {Ralf}\
			\bibnamefont {Eichhorn}}, \ and\ \bibinfo {author} {\bibfnamefont {Supriya}\
			\bibnamefont {Krishnamurthy}},\ }\bibfield  {title} {\enquote {\bibinfo
			{title} {Efficiency fluctuations in microscopic machines},}\ }\href {\doibase
		10.1103/PhysRevLett.122.140601} {\bibfield  {journal} {\bibinfo  {journal}
			{Phys. Rev. Lett.}\ }\textbf {\bibinfo {volume} {122}},\ \bibinfo {pages}
		{140601} (\bibinfo {year} {2019})}\BibitemShut {NoStop}%
	\bibitem [{\citenamefont {Esposito}\ \emph
		{et~al.}(2010{\natexlab{a}})\citenamefont {Esposito}, \citenamefont {Kawai},
		\citenamefont {Lindenberg},\ and\ \citenamefont {Van~den
			Broeck}}]{Esposito10}%
	\BibitemOpen
	\bibfield  {author} {\bibinfo {author} {\bibfnamefont {Massimiliano}\
			\bibnamefont {Esposito}}, \bibinfo {author} {\bibfnamefont {Ryoichi}\
			\bibnamefont {Kawai}}, \bibinfo {author} {\bibfnamefont {Katja}\ \bibnamefont
			{Lindenberg}}, \ and\ \bibinfo {author} {\bibfnamefont {Christian}\
			\bibnamefont {Van~den Broeck}},\ }\bibfield  {title} {\enquote {\bibinfo
			{title} {Quantum-dot carnot engine at maximum power},}\ }\href {\doibase
		10.1103/PhysRevE.81.041106} {\bibfield  {journal} {\bibinfo  {journal} {Phys.
				Rev. E}\ }\textbf {\bibinfo {volume} {81}},\ \bibinfo {pages} {041106}
		(\bibinfo {year} {2010}{\natexlab{a}})}\BibitemShut {NoStop}%
	\bibitem [{\citenamefont {Scully}\ \emph {et~al.}(2011)\citenamefont {Scully},
		\citenamefont {Chapin}, \citenamefont {Dorfman}, \citenamefont {Kim},\ and\
		\citenamefont {Svidzinsky}}]{Scully11}%
	\BibitemOpen
	\bibfield  {author} {\bibinfo {author} {\bibfnamefont {Marlan~O.}\
			\bibnamefont {Scully}}, \bibinfo {author} {\bibfnamefont {Kimberly~R.}\
			\bibnamefont {Chapin}}, \bibinfo {author} {\bibfnamefont {Konstantin~E.}\
			\bibnamefont {Dorfman}}, \bibinfo {author} {\bibfnamefont {Moochan~Barnabas}\
			\bibnamefont {Kim}}, \ and\ \bibinfo {author} {\bibfnamefont {Anatoly}\
			\bibnamefont {Svidzinsky}},\ }\bibfield  {title} {\enquote {\bibinfo {title}
			{Quantum heat engine power can be increased by noise-induced coherence},}\
	}\href {\doibase 10.1073/pnas.1110234108} {\bibfield  {journal} {\bibinfo
			{journal} {Proc. Natl. Acad. Sci.}\ }\textbf {\bibinfo {volume} {108}},\
		\bibinfo {pages} {15097--15100} (\bibinfo {year} {2011})}\BibitemShut
	{NoStop}%
	\bibitem [{\citenamefont {Campisi}\ and\ \citenamefont
		{Fazio}(2016)}]{Campisi16}%
	\BibitemOpen
	\bibfield  {author} {\bibinfo {author} {\bibfnamefont {Michele}\ \bibnamefont
			{Campisi}}\ and\ \bibinfo {author} {\bibfnamefont {Rosario}\ \bibnamefont
			{Fazio}},\ }\bibfield  {title} {\enquote {\bibinfo {title} {The power of a
				critical heat engine},}\ }\href {\doibase 10.1038/ncomms11895} {\bibfield
		{journal} {\bibinfo  {journal} {Nat. Commun.}\ }\textbf {\bibinfo {volume}
			{7}},\ \bibinfo {pages} {11895} (\bibinfo {year} {2016})}\BibitemShut
	{NoStop}%
	\bibitem [{\citenamefont {Holubec}\ and\ \citenamefont
		{Ryabov}(2016)}]{Holubec16a}%
	\BibitemOpen
	\bibfield  {author} {\bibinfo {author} {\bibfnamefont {Viktor}\ \bibnamefont
			{Holubec}}\ and\ \bibinfo {author} {\bibfnamefont {Artem}\ \bibnamefont
			{Ryabov}},\ }\bibfield  {title} {\enquote {\bibinfo {title} {Maximum
				efficiency of low-dissipation heat engines at arbitrary power},}\ }\href
	{\doibase 10.1088/1742-5468/2016/07/073204} {\bibfield  {journal} {\bibinfo
			{journal} {Journal of Statistical Mechanics: Theory and Experiment}\ }\textbf
		{\bibinfo {volume} {2016}},\ \bibinfo {pages} {073204} (\bibinfo {year}
		{2016})}\BibitemShut {NoStop}%
	\bibitem [{\citenamefont {Brandner}\ \emph {et~al.}(2017)\citenamefont
		{Brandner}, \citenamefont {Bauer},\ and\ \citenamefont
		{Seifert}}]{Brandner17}%
	\BibitemOpen
	\bibfield  {author} {\bibinfo {author} {\bibfnamefont {Kay}\ \bibnamefont
			{Brandner}}, \bibinfo {author} {\bibfnamefont {Michael}\ \bibnamefont
			{Bauer}}, \ and\ \bibinfo {author} {\bibfnamefont {Udo}\ \bibnamefont
			{Seifert}},\ }\bibfield  {title} {\enquote {\bibinfo {title} {Universal
				coherence-induced power losses of quantum heat engines in linear response},}\
	}\href {\doibase 10.1103/PhysRevLett.119.170602} {\bibfield  {journal}
		{\bibinfo  {journal} {Phys. Rev. Lett.}\ }\textbf {\bibinfo {volume} {119}},\
		\bibinfo {pages} {170602} (\bibinfo {year} {2017})}\BibitemShut {NoStop}%
	\bibitem [{\citenamefont {Denzler}\ and\ \citenamefont
		{Lutz}(2020)}]{Denzler20}%
	\BibitemOpen
	\bibfield  {author} {\bibinfo {author} {\bibfnamefont {Tobias}\ \bibnamefont
			{Denzler}}\ and\ \bibinfo {author} {\bibfnamefont {Eric}\ \bibnamefont
			{Lutz}},\ }\bibfield  {title} {\enquote {\bibinfo {title} {Power fluctuations
				in a finite-time quantum carnot engine},}\ }\href
	{https://arxiv.org/abs/2007.01034} {\bibfield  {journal} {\bibinfo  {journal}
			{arXiv:2007.01034}\ } (\bibinfo {year} {2020})}\BibitemShut {NoStop}%
	\bibitem [{\citenamefont {Abah}\ \emph {et~al.}(2012)\citenamefont {Abah},
		\citenamefont {Ro\ss{}nagel}, \citenamefont {Jacob}, \citenamefont {Deffner},
		\citenamefont {Schmidt-Kaler}, \citenamefont {Singer},\ and\ \citenamefont
		{Lutz}}]{Abah12}%
	\BibitemOpen
	\bibfield  {author} {\bibinfo {author} {\bibfnamefont {O.}~\bibnamefont
			{Abah}}, \bibinfo {author} {\bibfnamefont {J.}~\bibnamefont {Ro\ss{}nagel}},
		\bibinfo {author} {\bibfnamefont {G.}~\bibnamefont {Jacob}}, \bibinfo
		{author} {\bibfnamefont {S.}~\bibnamefont {Deffner}}, \bibinfo {author}
		{\bibfnamefont {F.}~\bibnamefont {Schmidt-Kaler}}, \bibinfo {author}
		{\bibfnamefont {K.}~\bibnamefont {Singer}}, \ and\ \bibinfo {author}
		{\bibfnamefont {E.}~\bibnamefont {Lutz}},\ }\bibfield  {title} {\enquote
		{\bibinfo {title} {Single-ion heat engine at maximum power},}\ }\href
	{\doibase 10.1103/PhysRevLett.109.203006} {\bibfield  {journal} {\bibinfo
			{journal} {Phys. Rev. Lett.}\ }\textbf {\bibinfo {volume} {109}},\ \bibinfo
		{pages} {203006} (\bibinfo {year} {2012})}\BibitemShut {NoStop}%
	\bibitem [{\citenamefont {Esposito}\ \emph
		{et~al.}(2010{\natexlab{b}})\citenamefont {Esposito}, \citenamefont {Kawai},
		\citenamefont {Lindenberg},\ and\ \citenamefont {Van~den
			Broeck}}]{Esposito10a}%
	\BibitemOpen
	\bibfield  {author} {\bibinfo {author} {\bibfnamefont {Massimiliano}\
			\bibnamefont {Esposito}}, \bibinfo {author} {\bibfnamefont {Ryoichi}\
			\bibnamefont {Kawai}}, \bibinfo {author} {\bibfnamefont {Katja}\ \bibnamefont
			{Lindenberg}}, \ and\ \bibinfo {author} {\bibfnamefont {Christian}\
			\bibnamefont {Van~den Broeck}},\ }\bibfield  {title} {\enquote {\bibinfo
			{title} {Efficiency at maximum power of low-dissipation carnot engines},}\
	}\href {\doibase 10.1103/PhysRevLett.105.150603} {\bibfield  {journal}
		{\bibinfo  {journal} {Phys. Rev. Lett.}\ }\textbf {\bibinfo {volume} {105}},\
		\bibinfo {pages} {150603} (\bibinfo {year} {2010}{\natexlab{b}})}\BibitemShut
	{NoStop}%
	\bibitem [{\citenamefont {Guo}\ \emph {et~al.}(2013)\citenamefont {Guo},
		\citenamefont {Wang}, \citenamefont {Wang},\ and\ \citenamefont
		{Chen}}]{Guo13}%
	\BibitemOpen
	\bibfield  {author} {\bibinfo {author} {\bibfnamefont {Juncheng}\
			\bibnamefont {Guo}}, \bibinfo {author} {\bibfnamefont {Junyi}\ \bibnamefont
			{Wang}}, \bibinfo {author} {\bibfnamefont {Yuan}\ \bibnamefont {Wang}}, \
		and\ \bibinfo {author} {\bibfnamefont {Jincan}\ \bibnamefont {Chen}},\
	}\bibfield  {title} {\enquote {\bibinfo {title} {Universal efficiency bounds
				of weak-dissipative thermodynamic cycles at the maximum power output},}\
	}\href {\doibase 10.1103/PhysRevE.87.012133} {\bibfield  {journal} {\bibinfo
			{journal} {Phys. Rev. E}\ }\textbf {\bibinfo {volume} {87}},\ \bibinfo
		{pages} {012133} (\bibinfo {year} {2013})}\BibitemShut {NoStop}%
	\bibitem [{\citenamefont {Ma}\ \emph {et~al.}(2018)\citenamefont {Ma},
		\citenamefont {Xu}, \citenamefont {Dong},\ and\ \citenamefont {Sun}}]{Ma18a}%
	\BibitemOpen
	\bibfield  {author} {\bibinfo {author} {\bibfnamefont {Yu-Han}\ \bibnamefont
			{Ma}}, \bibinfo {author} {\bibfnamefont {Dazhi}\ \bibnamefont {Xu}}, \bibinfo
		{author} {\bibfnamefont {Hui}\ \bibnamefont {Dong}}, \ and\ \bibinfo {author}
		{\bibfnamefont {Chang-Pu}\ \bibnamefont {Sun}},\ }\bibfield  {title}
	{\enquote {\bibinfo {title} {Universal constraint for efficiency and power of
				a low-dissipation heat engine},}\ }\href {\doibase
		10.1103/PhysRevE.98.042112} {\bibfield  {journal} {\bibinfo  {journal} {Phys.
				Rev. E}\ }\textbf {\bibinfo {volume} {98}},\ \bibinfo {pages} {042112}
		(\bibinfo {year} {2018})}\BibitemShut {NoStop}%
	\bibitem [{\citenamefont {Pietzonka}\ and\ \citenamefont
		{Seifert}(2018)}]{Pietzonka18}%
	\BibitemOpen
	\bibfield  {author} {\bibinfo {author} {\bibfnamefont {Patrick}\ \bibnamefont
			{Pietzonka}}\ and\ \bibinfo {author} {\bibfnamefont {Udo}\ \bibnamefont
			{Seifert}},\ }\bibfield  {title} {\enquote {\bibinfo {title} {Universal
				trade-off between power, efficiency, and constancy in steady-state heat
				engines},}\ }\href {\doibase 10.1103/PhysRevLett.120.190602} {\bibfield
		{journal} {\bibinfo  {journal} {Phys. Rev. Lett.}\ }\textbf {\bibinfo
			{volume} {120}},\ \bibinfo {pages} {190602} (\bibinfo {year}
		{2018})}\BibitemShut {NoStop}%
	\bibitem [{\citenamefont {Holubec}\ and\ \citenamefont
		{Ryabov}(2018)}]{Holubec18}%
	\BibitemOpen
	\bibfield  {author} {\bibinfo {author} {\bibfnamefont {Viktor}\ \bibnamefont
			{Holubec}}\ and\ \bibinfo {author} {\bibfnamefont {Artem}\ \bibnamefont
			{Ryabov}},\ }\bibfield  {title} {\enquote {\bibinfo {title} {Cycling tames
				power fluctuations near optimum efficiency},}\ }\href {\doibase
		10.1103/PhysRevLett.121.120601} {\bibfield  {journal} {\bibinfo  {journal}
			{Phys. Rev. Lett.}\ }\textbf {\bibinfo {volume} {121}},\ \bibinfo {pages}
		{120601} (\bibinfo {year} {2018})}\BibitemShut {NoStop}%
	\bibitem [{\citenamefont {Dorfman}\ \emph {et~al.}(2018)\citenamefont
		{Dorfman}, \citenamefont {Xu},\ and\ \citenamefont {Cao}}]{Dorfman18}%
	\BibitemOpen
	\bibfield  {author} {\bibinfo {author} {\bibfnamefont {Konstantin~E.}\
			\bibnamefont {Dorfman}}, \bibinfo {author} {\bibfnamefont {Dazhi}\
			\bibnamefont {Xu}}, \ and\ \bibinfo {author} {\bibfnamefont {Jianshu}\
			\bibnamefont {Cao}},\ }\bibfield  {title} {\enquote {\bibinfo {title}
			{Efficiency at maximum power of a laser quantum heat engine enhanced by
				noise-induced coherence},}\ }\href {\doibase 10.1103/PhysRevE.97.042120}
	{\bibfield  {journal} {\bibinfo  {journal} {Phys. Rev. E}\ }\textbf {\bibinfo
			{volume} {97}},\ \bibinfo {pages} {042120} (\bibinfo {year}
		{2018})}\BibitemShut {NoStop}%
	\bibitem [{\citenamefont {Abiuso}\ and\ \citenamefont
		{Perarnau-Llobet}(2020)}]{Abiuso20}%
	\BibitemOpen
	\bibfield  {author} {\bibinfo {author} {\bibfnamefont {Paolo}\ \bibnamefont
			{Abiuso}}\ and\ \bibinfo {author} {\bibfnamefont {Mart\'{\i}}\ \bibnamefont
			{Perarnau-Llobet}},\ }\bibfield  {title} {\enquote {\bibinfo {title} {Optimal
				cycles for low-dissipation heat engines},}\ }\href {\doibase
		10.1103/PhysRevLett.124.110606} {\bibfield  {journal} {\bibinfo  {journal}
			{Phys. Rev. Lett.}\ }\textbf {\bibinfo {volume} {124}},\ \bibinfo {pages}
		{110606} (\bibinfo {year} {2020})}\BibitemShut {NoStop}%
	\bibitem [{\citenamefont {Brandner}\ and\ \citenamefont
		{Saito}(2020)}]{Brandner20}%
	\BibitemOpen
	\bibfield  {author} {\bibinfo {author} {\bibfnamefont {Kay}\ \bibnamefont
			{Brandner}}\ and\ \bibinfo {author} {\bibfnamefont {Keiji}\ \bibnamefont
			{Saito}},\ }\bibfield  {title} {\enquote {\bibinfo {title} {Thermodynamic
				geometry of microscopic heat engines},}\ }\href {\doibase
		10.1103/PhysRevLett.124.040602} {\bibfield  {journal} {\bibinfo  {journal}
			{Phys. Rev. Lett.}\ }\textbf {\bibinfo {volume} {124}},\ \bibinfo {pages}
		{040602} (\bibinfo {year} {2020})}\BibitemShut {NoStop}%
	\bibitem [{\citenamefont {Miller}\ and\ \citenamefont
		{Mehboudi}(2020)}]{Miller20}%
	\BibitemOpen
	\bibfield  {author} {\bibinfo {author} {\bibfnamefont {Harry J.~D.}\
			\bibnamefont {Miller}}\ and\ \bibinfo {author} {\bibfnamefont {Mohammad}\
			\bibnamefont {Mehboudi}},\ }\bibfield  {title} {\enquote {\bibinfo {title}
			{Geometry of work fluctuations versus efficiency in microscopic thermal
				machines},}\ }\href {\doibase 10.1103/PhysRevLett.125.260602} {\bibfield
		{journal} {\bibinfo  {journal} {Phys. Rev. Lett.}\ }\textbf {\bibinfo
			{volume} {125}},\ \bibinfo {pages} {260602} (\bibinfo {year}
		{2020})}\BibitemShut {NoStop}%
	\bibitem [{\citenamefont {Singh}(2020)}]{Singh20}%
	\BibitemOpen
	\bibfield  {author} {\bibinfo {author} {\bibfnamefont {Varinder}\
			\bibnamefont {Singh}},\ }\bibfield  {title} {\enquote {\bibinfo {title}
			{Optimal operation of a three-level quantum heat engine and universal nature
				of efficiency},}\ }\href {\doibase 10.1103/PhysRevResearch.2.043187}
	{\bibfield  {journal} {\bibinfo  {journal} {Phys. Rev. Research}\ }\textbf
		{\bibinfo {volume} {2}},\ \bibinfo {pages} {043187} (\bibinfo {year}
		{2020})}\BibitemShut {NoStop}%
	\bibitem [{\citenamefont {Benenti}\ \emph {et~al.}(2020)\citenamefont
		{Benenti}, \citenamefont {Casati},\ and\ \citenamefont {Wang}}]{Benenti20}%
	\BibitemOpen
	\bibfield  {author} {\bibinfo {author} {\bibfnamefont {Giuliano}\
			\bibnamefont {Benenti}}, \bibinfo {author} {\bibfnamefont {Giulio}\
			\bibnamefont {Casati}}, \ and\ \bibinfo {author} {\bibfnamefont {Jiao}\
			\bibnamefont {Wang}},\ }\bibfield  {title} {\enquote {\bibinfo {title}
			{Power, efficiency, and fluctuations in steady-state heat engines},}\ }\href
	{\doibase 10.1103/PhysRevE.102.040103} {\bibfield  {journal} {\bibinfo
			{journal} {Phys. Rev. E}\ }\textbf {\bibinfo {volume} {102}},\ \bibinfo
		{pages} {040103} (\bibinfo {year} {2020})}\BibitemShut {NoStop}%
	\bibitem [{\citenamefont {Ro{\ss}nagel}\ \emph {et~al.}(2016)\citenamefont
		{Ro{\ss}nagel}, \citenamefont {Dawkins}, \citenamefont {Tolazzi},
		\citenamefont {Abah}, \citenamefont {Lutz}, \citenamefont {Schmidt-Kaler},\
		and\ \citenamefont {Singer}}]{Rossnagel16}%
	\BibitemOpen
	\bibfield  {author} {\bibinfo {author} {\bibfnamefont {Johannes}\
			\bibnamefont {Ro{\ss}nagel}}, \bibinfo {author} {\bibfnamefont {Samuel~T.}\
			\bibnamefont {Dawkins}}, \bibinfo {author} {\bibfnamefont {Karl~N.}\
			\bibnamefont {Tolazzi}}, \bibinfo {author} {\bibfnamefont {Obinna}\
			\bibnamefont {Abah}}, \bibinfo {author} {\bibfnamefont {Eric}\ \bibnamefont
			{Lutz}}, \bibinfo {author} {\bibfnamefont {Ferdinand}\ \bibnamefont
			{Schmidt-Kaler}}, \ and\ \bibinfo {author} {\bibfnamefont {Kilian}\
			\bibnamefont {Singer}},\ }\bibfield  {title} {\enquote {\bibinfo {title} {A
				single-atom heat engine},}\ }\href {\doibase 10.1126/science.aad6320}
	{\bibfield  {journal} {\bibinfo  {journal} {Science}\ }\textbf {\bibinfo
			{volume} {352}},\ \bibinfo {pages} {325--329} (\bibinfo {year}
		{2016})}\BibitemShut {NoStop}%
	\bibitem [{\citenamefont {Saryal}\ and\ \citenamefont
		{Agarwalla}(2021)}]{Saryal2021}%
	\BibitemOpen
	\bibfield  {author} {\bibinfo {author} {\bibfnamefont {Sushant}\ \bibnamefont
			{Saryal}}\ and\ \bibinfo {author} {\bibfnamefont {Bijay~Kumar}\ \bibnamefont
			{Agarwalla}},\ }\bibfield  {title} {\enquote {\bibinfo {title} {Bounds on
				fluctuations for finite-time quantum otto cycle},}\ }\href
	{https://arxiv.org/abs/2104.12173} {\bibfield  {journal} {\bibinfo  {journal}
			{arXiv:2104.12173}\ } (\bibinfo {year} {2021})}\BibitemShut {NoStop}%
	\bibitem [{\citenamefont {Saryal}\ \emph {et~al.}(2021)\citenamefont {Saryal},
		\citenamefont {Gerry}, \citenamefont {Khait}, \citenamefont {Segal},\ and\
		\citenamefont {Agarwalla}}]{Saryal2021a}%
	\BibitemOpen
	\bibfield  {author} {\bibinfo {author} {\bibfnamefont {Sushant}\ \bibnamefont
			{Saryal}}, \bibinfo {author} {\bibfnamefont {Matthew}\ \bibnamefont {Gerry}},
		\bibinfo {author} {\bibfnamefont {Ilia}\ \bibnamefont {Khait}}, \bibinfo
		{author} {\bibfnamefont {Dvira}\ \bibnamefont {Segal}}, \ and\ \bibinfo
		{author} {\bibfnamefont {Bijay~Kumar}\ \bibnamefont {Agarwalla}},\ }\bibfield
	{title} {\enquote {\bibinfo {title} {Universal bounds on fluctuations in
				continuous thermal machines},}\ }\href {https://arxiv.org/abs/2103.13513}
	{\bibfield  {journal} {\bibinfo  {journal} {arXiv:2103.13513}\ } (\bibinfo
		{year} {2021})}\BibitemShut {NoStop}%
	\bibitem [{\citenamefont {Bera}\ \emph {et~al.}(2021)\citenamefont {Bera},
		\citenamefont {Lewenstein},\ and\ \citenamefont {Bera}}]{MLBera21}%
	\BibitemOpen
	\bibfield  {author} {\bibinfo {author} {\bibfnamefont {Mohit~Lal}\
			\bibnamefont {Bera}}, \bibinfo {author} {\bibfnamefont {Maciej}\ \bibnamefont
			{Lewenstein}}, \ and\ \bibinfo {author} {\bibfnamefont {Manabendra~Nath}\
			\bibnamefont {Bera}},\ }\bibfield  {title} {\enquote {\bibinfo {title}
			{Attaining {C}arnot efficiency with quantum and nanoscale heat engines},}\
	}\href {\doibase 10.1038/s41534-021-00366-6} {\bibfield  {journal} {\bibinfo
			{journal} {npj Quantum Information}\ }\textbf {\bibinfo {volume} {7}},\
		\bibinfo {pages} {31} (\bibinfo {year} {2021})}\BibitemShut {NoStop}%
	\bibitem [{\citenamefont {Anandan}\ and\ \citenamefont
		{Aharonov}(1990)}]{Anandan90}%
	\BibitemOpen
	\bibfield  {author} {\bibinfo {author} {\bibfnamefont {J.}~\bibnamefont
			{Anandan}}\ and\ \bibinfo {author} {\bibfnamefont {Y.}~\bibnamefont
			{Aharonov}},\ }\bibfield  {title} {\enquote {\bibinfo {title} {Geometry of
				quantum evolution},}\ }\href {\doibase 10.1103/PhysRevLett.65.1697}
	{\bibfield  {journal} {\bibinfo  {journal} {Phys. Rev. Lett.}\ }\textbf
		{\bibinfo {volume} {65}},\ \bibinfo {pages} {1697--1700} (\bibinfo {year}
		{1990})}\BibitemShut {NoStop}%
	\bibitem [{\citenamefont {Gerry}\ and\ \citenamefont {Eberly}(1990)}]{Gerry90}%
	\BibitemOpen
	\bibfield  {author} {\bibinfo {author} {\bibfnamefont {Christopher~C.}\
			\bibnamefont {Gerry}}\ and\ \bibinfo {author} {\bibfnamefont {J.~H.}\
			\bibnamefont {Eberly}},\ }\bibfield  {title} {\enquote {\bibinfo {title}
			{Dynamics of a raman coupled model interacting with two quantized cavity
				fields},}\ }\href {\doibase 10.1103/PhysRevA.42.6805} {\bibfield  {journal}
		{\bibinfo  {journal} {Phys. Rev. A}\ }\textbf {\bibinfo {volume} {42}},\
		\bibinfo {pages} {6805--6815} (\bibinfo {year} {1990})}\BibitemShut {NoStop}%
	\bibitem [{\citenamefont {Wu}(1996)}]{Wu96}%
	\BibitemOpen
	\bibfield  {author} {\bibinfo {author} {\bibfnamefont {Ying}\ \bibnamefont
			{Wu}},\ }\bibfield  {title} {\enquote {\bibinfo {title} {Effective raman
				theory for a three-level atom in the \ensuremath{\Lambda} configuration},}\
	}\href {\doibase 10.1103/PhysRevA.54.1586} {\bibfield  {journal} {\bibinfo
			{journal} {Phys. Rev. A}\ }\textbf {\bibinfo {volume} {54}},\ \bibinfo
		{pages} {1586--1592} (\bibinfo {year} {1996})}\BibitemShut {NoStop}%
	\bibitem [{\citenamefont {Greentree}\ \emph {et~al.}(2013)\citenamefont
		{Greentree}, \citenamefont {Koch},\ and\ \citenamefont
		{Larson}}]{Greentree2013}%
	\BibitemOpen
	\bibfield  {author} {\bibinfo {author} {\bibfnamefont {Andrew~D}\
			\bibnamefont {Greentree}}, \bibinfo {author} {\bibfnamefont {Jens}\
			\bibnamefont {Koch}}, \ and\ \bibinfo {author} {\bibfnamefont {Jonas}\
			\bibnamefont {Larson}},\ }\bibfield  {title} {\enquote {\bibinfo {title}
			{Fifty years of jaynes{\textendash}cummings physics},}\ }\href {\doibase
		10.1088/0953-4075/46/22/220201} {\bibfield  {journal} {\bibinfo  {journal}
			{J. Phys. B: At. Mol. Opt. Phys.}\ }\textbf {\bibinfo {volume} {46}},\
		\bibinfo {pages} {220201} (\bibinfo {year} {2013})}\BibitemShut {NoStop}%
	\bibitem [{\citenamefont {Scovil}\ and\ \citenamefont
		{Schulz-DuBois}(1959)}]{Scovil59}%
	\BibitemOpen
	\bibfield  {author} {\bibinfo {author} {\bibfnamefont {H.~E.~D.}\
			\bibnamefont {Scovil}}\ and\ \bibinfo {author} {\bibfnamefont {E.~O.}\
			\bibnamefont {Schulz-DuBois}},\ }\bibfield  {title} {\enquote {\bibinfo
			{title} {Three-level masers as heat engines},}\ }\href {\doibase
		10.1103/PhysRevLett.2.262} {\bibfield  {journal} {\bibinfo  {journal} {Phys.
				Rev. Lett.}\ }\textbf {\bibinfo {volume} {2}},\ \bibinfo {pages} {262--263}
		(\bibinfo {year} {1959})}\BibitemShut {NoStop}%
	\bibitem [{\citenamefont {Ghosh}\ \emph {et~al.}(2018)\citenamefont {Ghosh},
		\citenamefont {Gelbwaser-Klimovsky}, \citenamefont {Niedenzu}, \citenamefont
		{Lvovsky}, \citenamefont {Mazets}, \citenamefont {Scully},\ and\
		\citenamefont {Kurizki}}]{Ghosh18}%
	\BibitemOpen
	\bibfield  {author} {\bibinfo {author} {\bibfnamefont {Arnab}\ \bibnamefont
			{Ghosh}}, \bibinfo {author} {\bibfnamefont {David}\ \bibnamefont
			{Gelbwaser-Klimovsky}}, \bibinfo {author} {\bibfnamefont {Wolfgang}\
			\bibnamefont {Niedenzu}}, \bibinfo {author} {\bibfnamefont {Alexander~I.}\
			\bibnamefont {Lvovsky}}, \bibinfo {author} {\bibfnamefont {Igor}\
			\bibnamefont {Mazets}}, \bibinfo {author} {\bibfnamefont {Marlan~O.}\
			\bibnamefont {Scully}}, \ and\ \bibinfo {author} {\bibfnamefont {Gershon}\
			\bibnamefont {Kurizki}},\ }\bibfield  {title} {\enquote {\bibinfo {title}
			{Two-level masers as heat-to-work converters},}\ }\href {\doibase
		10.1073/pnas.1805354115} {\bibfield  {journal} {\bibinfo  {journal} {Proc.
				Natl. Acad. Sci.}\ }\textbf {\bibinfo {volume} {115}},\ \bibinfo {pages}
		{9941--9944} (\bibinfo {year} {2018})}\BibitemShut {NoStop}%
\end{thebibliography}
%

\end{document}